%% file: verneuil.tex
\documentclass[a4paper,12pt]{article}
\usepackage{graphicx}
\usepackage{axodraw}
\begin{document}

\def\brique{}
\def\Black{}
\def\noir{}
\def\Blue{}
\def\Green{}
\def\rouge{}
\def\bleu{}
\def\vert{}
\def\violet{}

\newcommand{\be}{\begin{equation}}
\newcommand{\beq}{\begin{equation}}
\newcommand{\eeq}{\end{equation}}
\newcommand{\ee}{\end{equation}}

\newcommand{\beqn}{\begin{eqnarray}}
\newcommand{\eeqn}{\end{eqnarray}}
\newcommand{\bea}{\begin{eqnarray}}
\newcommand{\eea}{\end{eqnarray}}
\newcommand{\ena}{\end{eqnarray}}
\newcommand{\ra}{\rightarrow}
\newcommand{\susy}{{{\cal SUSY}$\;$}}
\newcommand{\su}{$ SU(2) \times U(1)\,$}

\newcommand{\gag}{$\gamma \gamma$ }
\newcommand{\gagt}{\gamma \gamma }
\newcommand{\gam}{\gamma \gamma }
\def\W{{\mbox{\boldmath $W$}}}
\def\B{{\mbox{\boldmath $B$}}}
\def\V{{\mbox{\boldmath $V$}}}
\newcommand{\np}{Nucl.\,Phys.\,}
\newcommand{\pl}{Phys.\,Lett.\,}
\newcommand{\pr}{Phys.\,Rev.\,}
\newcommand{\prl}{Phys.\,Rev.\,Lett.\,}
\newcommand{\prep}{Phys.\,Rep.\,}
\newcommand{\zp}{Z.\,Phys.\,}
\newcommand{\sovjnp}{{\em Sov.\ J.\ Nucl.\ Phys.\ }}
\newcommand{\nuclinst}{{\em Nucl.\ Instrum.\ Meth.\ }}
\newcommand{\annp}{{\em Ann.\ Phys.\ }}
\newcommand{\intjmp}{{\em Int.\ J.\ of Mod.\  Phys.\ }}

\newcommand{\eps}{\epsilon}
\newcommand{\mw}{M_{W}}
\newcommand{\mww}{M_{W}^{2}}
\newcommand{\mwmw}{M_{W}^{2}}
\newcommand{\mhmh}{M_{H}^2}
\newcommand{\mz}{M_{Z}}
\newcommand{\mzz}{M_{Z}^{2}}

\newcommand{\cw}{\cos\theta_W}
\newcommand{\sw}{\sin\theta_W}
\newcommand{\tw}{\tan\theta_W}
\def\cww{\cos^2\theta_W}
\def\sww{\sin^2\theta_W}
\def\tww{\tan^2\theta_W}

\newcommand{\epm}{$e^{+} e^{-}\;$}
\newcommand{\epemt}{$e^{+} e^{-}\;$}
\newcommand{\epem}{e^{+} e^{-}\;}
\newcommand{\ememt}{$e^{-} e^{-}\;$}
\newcommand{\emem}{e^{-} e^{-}\;}

\newcommand{\lra}{\leftrightarrow}
\newcommand{\tr}{{\rm Tr}}
\def\ls1{{\not l}_1}
\newcommand{\cms}{centre-of-mass\hspace*{.1cm}}


\newcommand{\dkg}{\Delta \kappa_{\gamma}}
\newcommand{\dkz}{\Delta \kappa_{Z}}
\newcommand{\dz}{\delta_{Z}}
\newcommand{\dgz}{\Delta g^{1}_{Z}}
\newcommand{\dgzt}{$\Delta g^{1}_{Z}\;$}
\newcommand{\la}{\lambda}
\newcommand{\lag}{\lambda_{\gamma}}
\newcommand{\lambdae}{\lambda_{e}}
\newcommand{\laz}{\lambda_{Z}}
\newcommand{\lnl}{L_{9L}}
\newcommand{\lnr}{L_{9R}}
\newcommand{\lt}{L_{10}}
\newcommand{\lu}{L_{1}}
\newcommand{\ld}{L_{2}}
\newcommand{\eeww}{e^{+} e^{-} \ra W^+ W^- \;}
\newcommand{\eewwt}{$e^{+} e^{-} \ra W^+ W^- \;$}
\newcommand{\epemww}{e^{+} e^{-} \ra W^+ W^- }
\newcommand{\epemwwt}{$e^{+} e^{-} \ra W^+ W^- \;$}
\newcommand{\eennhht}{$e^{+} e^{-} \ra \nu_e \bar \nu_e HH\;$}
\newcommand{\eennhh}{e^{+} e^{-} \ra \nu_e \bar \nu_e HH\;}
\newcommand{\ppwg}{p p \ra W \gamma}
\newcommand{\wwhh}{W^+ W^- \ra HH\;}
\newcommand{\wwhht}{$W^+ W^- \ra HH\;$}
\newcommand{\ppwz}{pp \ra W Z}
\newcommand{\ppwgt}{$p p \ra W \gamma \;$}
\newcommand{\ppwzt}{$pp \ra W Z \;$}
\newcommand{\gamgamt}{$\gamma \gamma \;$}
\newcommand{\gamgam}{\gamma \gamma \;}
\newcommand{\egamt}{$e \gamma \;$}
\newcommand{\egam}{e \gamma \;}
\newcommand{\gamgamwwt}{$\gamma \gamma \ra W^+ W^- \;$}
\newcommand{\gamgamwwht}{$\gamma \gamma \ra W^+ W^- H \;$}
\newcommand{\gamgamwwh}{\gamma \gamma \ra W^+ W^- H \;}
\newcommand{\gamgamwwhht}{$\gamma \gamma \ra W^+ W^- H H\;$}
\newcommand{\gamgamwwhh}{\gamma \gamma \ra W^+ W^- H H\;}
\newcommand{\ggww}{\gamma \gamma \ra W^+ W^-}
\newcommand{\ggwwt}{$\gamma \gamma \ra W^+ W^- \;$}
\newcommand{\ggwwht}{$\gamma \gamma \ra W^+ W^- H \;$}
\newcommand{\ggwwh}{\gamma \gamma \ra W^+ W^- H \;}
\newcommand{\ggwwhht}{$\gamma \gamma \ra W^+ W^- H H\;$}
\newcommand{\ggwwhh}{\gamma \gamma \ra W^+ W^- H H\;}
\newcommand{\ggwwz}{\gamma \gamma \ra W^+ W^- Z\;}
\newcommand{\ggwwzt}{$\gamma \gamma \ra W^+ W^- Z\;$}

\newcommand{\ptu}{p_{1\bot}}
\newcommand{\vecptu}{\vec{p}_{1\bot}}
\newcommand{\ptd}{p_{2\bot}}
\newcommand{\vecptd}{\vec{p}_{2\bot}}
\newcommand{\ie}{{\em i.e.}}
\newcommand{\cm}{{{\cal M}}}
\newcommand{\cl}{{{\cal L}}}
\newcommand{\cd}{{{\cal D}}}
\newcommand{\cv}{{{\cal V}}}
\def\slashc{c\kern -.400em {/}}
\def\slashp{p\kern -.400em {/}}
\def\slashL{L\kern -.450em {/}}
\def\slashcl{\cl\kern -.600em {/}}
\def\slashD{D\kern -.600em {/}}
\def\Ww{{\mbox{\boldmath $W$}}}
\def\B{{\mbox{\boldmath $B$}}}
\def\noi{\noindent}
\def\nn{\noindent}
\def\sm{${\cal{S}} {\cal{M}}\;$}
\def\smn{${\cal{S}} {\cal{M}}$}
\def\nph{${\cal{N}} {\cal{P}}\;$}
\def\sb{$ {\cal{S}}  {\cal{B}}\;$}
\def\ssb{${\cal{S}} {\cal{S}}  {\cal{B}}\;$}
\def\ssbe{{\cal{S}} {\cal{S}}  {\cal{B}}}
\def\cviol{${\cal{C}}\;$}
\def\pviol{${\cal{P}}\;$}
\def\cpviol{${\cal{C}} {\cal{P}}\;$}

\newcommand{\lgg}{\lambda_1\lambda_2}
\newcommand{\lww}{\lambda_3\lambda_4}
\newcommand{\ppin}{ P^+_{12}}
\newcommand{\pmin}{ P^-_{12}}
\newcommand{\ppout}{ P^+_{34}}
\newcommand{\pmout}{ P^-_{34}}
\newcommand{\sinsq}{\sin^2\theta}
\newcommand{\cossq}{\cos^2\theta}
\newcommand{\yt}{y_\theta}
\newcommand{\hppll}{++;00}
\newcommand{\hpmll}{+-;00}
\newcommand{\hpplt}{++;\lambda_30}
\newcommand{\hpmlt}{+-;\lambda_30}
\newcommand{\hpptt}{++;\lambda_3\lambda_4}
\newcommand{\hpmtt}{+-;\lambda_3\lambda_4}
\newcommand{\dk}{\Delta\kappa}
\newcommand{\klam}{\Delta\kappa \lambda_\gamma }
\newcommand{\kac}{\Delta\kappa^2 }
\newcommand{\lac}{\lambda_\gamma^2 }
\def\gamgamtzz{$\gamma \gamma \ra ZZ \;$}
\def\gamgamtww{$\gamma \gamma \ra W^+ W^-\;$}
\def\gamgamtwwe{\gamma \gamma \ra W^+ W^-}
\def\sinb{\sin\beta}
\def\cosb{\cos\beta}
\def\sinbb{\sin (2\beta)}
\def\cosbb{\cos (2 \beta)}
\def\tgb{\tan \beta}
\def\tgbt{$\tan \beta\;\;$}
\def\tgbsq{\tan^2 \beta}
\def\sinal{\sin\alpha}
\def\cosal{\cos\alpha}
\def\stop{\tilde{t}}
\def\sto{\tilde{t}_1}
\def\stt{\tilde{t}_2}
\def\stl{\tilde{t}_L}
\def\str{\tilde{t}_R}
\def\msto{m_{\sto}}
\def\mstosq{m_{\sto}^2}
\def\mstt{m_{\stt}}
\def\msttsq{m_{\stt}^2}
\def\mt{m_t}
\def\mtsq{m_t^2}
\def\sint{\sin\theta_{\stop}}
\def\sintt{\sin 2\theta_{\stop}}
\def\cost{\cos\theta_{\stop}}
\def\sintsq{\sin^2\theta_{\stop}}
\def\costsq{\cos^2\theta_{\stop}}
\def\mqtt{\M_{\tilde{Q}_3}^2}
\def\mutt{\M_{\tilde{U}_{3R}}^2}
\def\sbottom{\tilde{b}}
\def\sbo{\tilde{b}_1}
\def\sbt{\tilde{b}_2}
\def\sbl{\tilde{b}_L}
\def\sbr{\tilde{b}_R}
\def\msbo{m_{\sbo}}
\def\msbosq{m_{\sbo}^2}
\def\msbt{m_{\sbt}}
\def\msbtsq{m_{\sbt}^2}
\def\mt{m_t}
\def\mtsq{m_t^2}
\def\selectron{\tilde{e}}
\def\seo{\tilde{e}_1}
\def\set{\tilde{e}_2}
\def\sel{\tilde{e}_L}
\def\ser{\tilde{e}_R}
\def\mseo{m_{\seo}}
\def\mseosq{m_{\seo}^2}
\def\mset{m_{\set}}
\def\msetsq{m_{\set}^2}
\def\msel{m_{\sel}}
\def\mser{m_{\ser}}
\def\me{m_e}
\def\mesq{m_e^2}
\def\snu{\tilde{\nu}}
\def\snue{\tilde{\nu_e}}
\def\set{\tilde{e}_2}
\def\snul{\tilde{\nu}_L}
\def\msnue{m_{\snue}}
\def\msnuesq{m_{\snue}^2}
\def\smuon{\tilde{\mu}}
\def\smul{\tilde{\mu}_L}
\def\smur{\tilde{\mu}_R}
\def\msmul{m_{\smul}}
\def\msmulsq{m_{\smul}^2}
\def\msmur{m_{\smur}}
\def\msmursq{m_{\smur}^2}
\def\stau{\tilde{\tau}}
\def\stauo{\tilde{\tau}_1}
\def\staut{\tilde{\tau}_2}
\def\staul{\tilde{\tau}_L}
\def\staur{\tilde{\tau}_R}
\def\mstauo{m_{\stauo}}
\def\mstauosq{m_{\stauo}^2}
\def\mstaut{m_{\staut}}
\def\mstautsq{m_{\staut}^2}
\def\mtau{m_\tau}
\def\mtausq{m_\tau^2}
\def\gluino{\tilde{g}}
\def\mgluino{m_{\tilde{g}}}
\def\mchi{m_\chi^+}
\def\neuto{\tilde{\chi}_1^0}
\def\mneuto{m_{\tilde{\chi}_1^0}}
\def\neutt{\tilde{\chi}_2^0}
\def\mneutt{m_{\tilde{\chi}_2^0}}
\def\neutth{\tilde{\chi}_3^0}
\def\mneutth{m_{\tilde{\chi}_3^0}}
\def\neutf{\tilde{\chi}_4^0}
\def\mneutf{m_{\tilde{\chi}_4^0}}
\def\chargop{\tilde{\chi}_1^+}
\def\mchargo{m_{\tilde{\chi}_1^+}}
\def\chargtp{\tilde{\chi}_2^+}
\def\mchargt{m_{\tilde{\chi}_2^+}}
\def\chargom{\tilde{\chi}_1^-}
\def\chargtm{\tilde{\chi}_2^-}
\def\bino{\tilde{b}}
\def\wino{\tilde{w}}
\def\photino{\tilde{\gamma}}
\def\zino{tilde{z}}
\def\sdowno{\tilde{d}_1}
\def\sdownt{\tilde{d}_2}
\def\sdownl{\tilde{d}_L}
\def\sdownr{\tilde{d}_R}
\def\supo{\tilde{u}_1}
\def\supt{\tilde{u}_2}
\def\supl{\tilde{u}_L}
\def\supr{\tilde{u}_R}
\def\mh{m_h}
\def\mht{m_h^2}
\def\MH{M_H}
\def\MHt{M_H^2}
\def\MA{M_A}
\def\MAt{M_A^2}
\def\MHp{M_H^+}
\def\MHm{M_H^-}

\noir

\begin{titlepage}
\def\baselinestretch{1.2}
\topmargin     -0.25in


\vspace*{\fill}
\begin{center}
{\large {\bf  Physics for the future colliders.}}

\vspace*{0.5cm}

\begin{tabular}[t]{c}

{\bf  Fawzi Boudjema }
 \\
\\
\\
{\it  Laboratoire de Physique Th\'eorique} {\large LAPTH}
\footnote{URA 14-36 du CNRS, associ\'ee  \`a
l'Universit\'e de Savoie.}\\
 {\it Chemin de Bellevue, B.P. 110, F-74941 Annecy-le-Vieux,
Cedex, France.}\\
\end{tabular}
\end{center}

\centerline{ {\bf Abstract} } \baselineskip=14pt \noindent
 {\small Physics issues at the upcoming and planned colliders are discussed. We critically review the
 the different arguments that suggest that New Physics is bound to
 materialise at the TeV scale and why we should keep an open
 minded approach. The complementarity of the LHC and a moderate
 energy \epemt collider is stressed together with the need for
 higher energy machines
.}
\vspace*{\fill}

Talk given at the Meeting "Which Colliders for the Future?",
Orsay, Paris, Sep. 2000.

\vspace*{0.1cm} \rightline{LAPTH-Conf837/01}
\end{titlepage}

\section{Introduction: The particles of today}
The extremely successful Standard Model, \smn, is a theory that
describes in a very neat, and in part economical, manner the
interactions of all known fundamental spin-1 and spin-$1/2$
particles. However it also requires  a spin-0 particle which has
so far not been seen. Future experiments are needed either to
unravel this missing particle or probe whether the scalar sector
is in fact the tip of a beautiful iceberg full of much interesting
physics to come. When trying to predict what physics is to be
expected at high energies and which machines are best suitable for
this physics one should ideally rid oneself of all theoretical
prejudices even though some theoretical arguments backed by some
very strong indirect experimental
evidence are very tempting.\\

\begin{figure*}[htbp]
\begin{center}
\mbox{
\includegraphics[width=8cm,height=8cm]{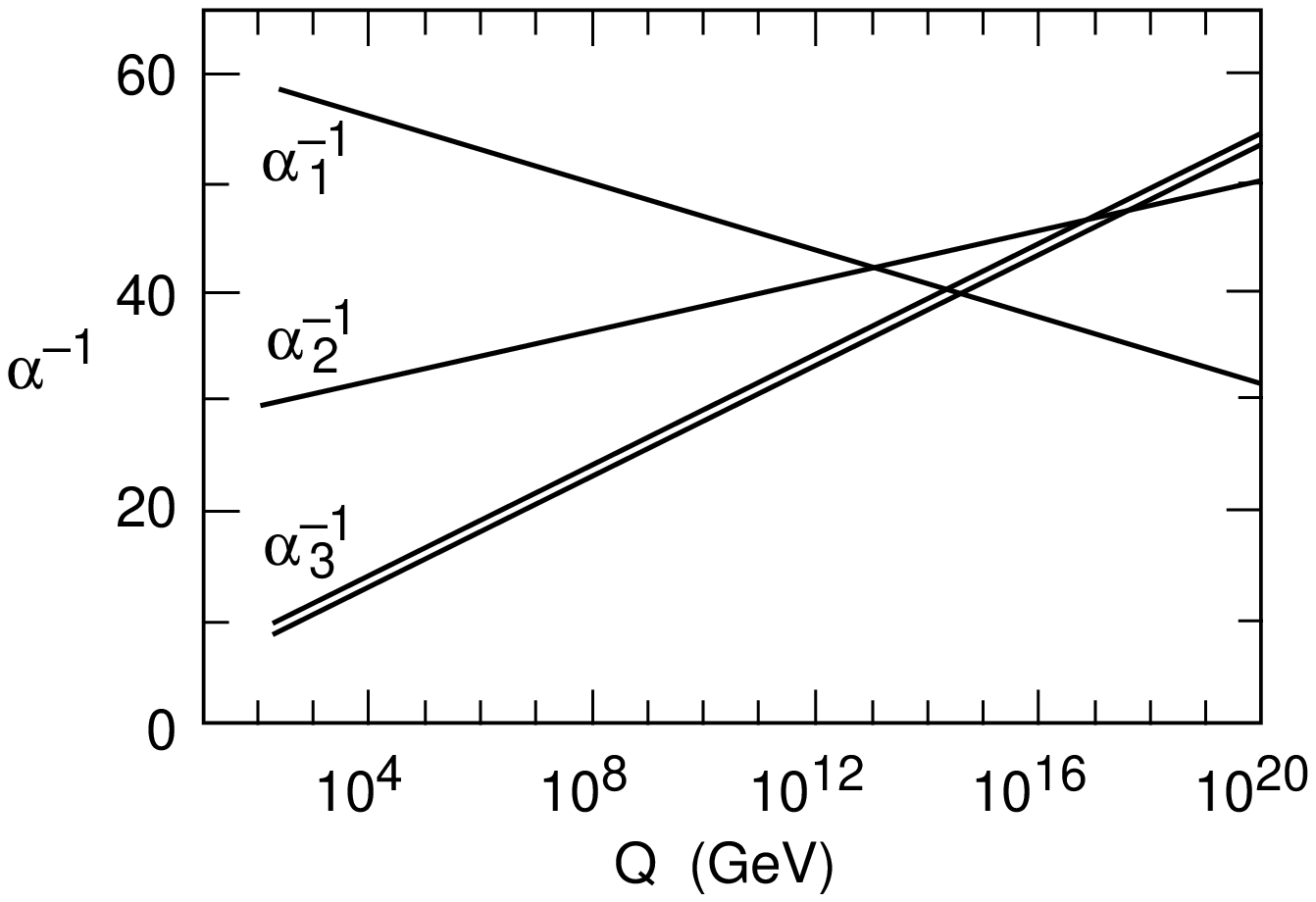}
\includegraphics[width=8cm,height=8cm]{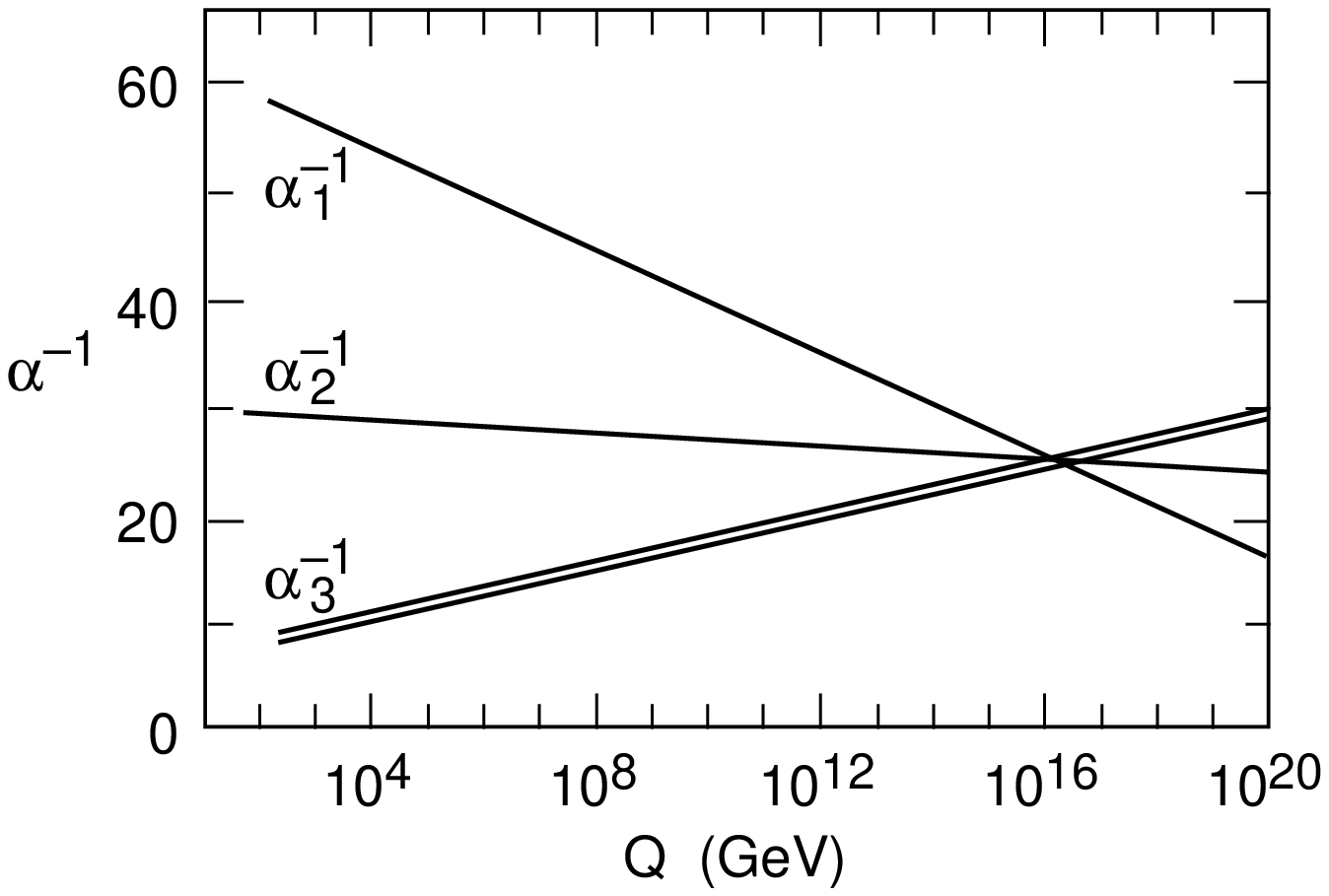}}
\caption{\label{gaugeuni}{\em Evolution of the gauge couplings in
the \sm and in SUSY with particle masses at a TeV scale, from
\cite{Peskingaugeuni} \/.}}
\end{center}
\end{figure*}

Take for instance the {\bf spin-$1$ sector}. LEP has brought a
beyond-doubt confirmation that all the forces of the theory are
based on a local gauge symmetry, by the fact that all the
couplings of the gauge bosons not only to fermions but among
themselves are universal. Moreover the three independent gauge
couplings for the $SU(2)\times U(1) \times SU(3)$ forces are now
measured with such a precision that evolving them to high scale
seems  to suggest that they might unify at a scale of $\sim 2\;
10^{16}$GeV, if one postulates the existence of supersymmetric
particles at energies of the order $1$TeV, or even below, see
Figs~\ref{gaugeuni}. Since many of the problems that plague the
Standard Model are solved in the more symmetric supersymmetric
completion of the model, many view this unification as strong
evidence for SUSY and that SUSY will be discovered at the LHC. To
be cautious, one should also point out that, recently, one has
found out\cite{Dudasuni} that models  based on extra-dimensions,
not necessarily supersymmetric, are also able to unify the three
gauge couplings at scales closer to the TeV scale than the Planck
scale, without guarantee that new particles be seen at the
upcoming colliders. When talking about the forces, one should not
forget that the all pervading gravity still lacks a satisfactory
quantum description. Although it is not included as part of the
\sm, it does provide a scale, the Planck scale $M_P$, which
conceptually
poses a severe conflict with the scale of the electroweak theory. \\
{\bf The spin-$1/2$}. With the relatively recent discovery of the
top and the direct confirmation of the tau neutrino the matter
content of the standard model is complete. The Lagrangian
describing the interaction of the spin-$1/2$ and spin-$1$ of the
theory, leaving aside the masses, is very simple and economical
\beqn
{{\cal L}}_{1,1/2}= \;i\;\bar{f} \slashD f -\frac{1}{4}
{\left(F_{\mu \nu}^i\right)}^2
\eeqn
with $D$ being the appropriate covariant derivative acting on the
left handed doublets and the right-handed singlet. The latter do
not mix in the absence of mass. Having defined the charges one
only needs the three coupling constants.

Let us now turn to the {\bf spin-$0$ part}. First,  as mentioned
earlier the Higgs, the physical spin-$0$ particle of the model,
has not been {\em confirmed}. However a scalar doublet, with a
non-zero expectation value is required in the \sm to provide mass
to all the particles in a gauge invariant way. It is here that the
majority of the seemingly haphazard parameters of the \sm are
hidden. Moreover it is important that one checks the structure of
this sector and the nature of the scalar potential. This is a
large part of the physics of the future. It can be argued that it
is not completely correct to claim that we have not seen any
spin-$0$ in the \sm. Indeed, the Goldstone bosons $\varphi$ of the
scalar doublet $\Phi$, are somehow the longitudinal modes of the
$W$ and $Z$
$$\Phi= \left(\begin{array}{c} \varphi^+ \\ \frac{1}{\sqrt{2}}(v+\rouge H\noir +i\varphi_3)
\end{array}\right) $$
and therefore it is only \rouge $H$ \noir that has not been seen
yet.

The scalar potential is also mysterious
\beqn
{\cal V}_{SSB}=\lambda \left[ \Phi^\dagger \Phi\ - \frac{v^2}{2}
\right]^2=\lambda \left[ \Phi^\dagger \Phi\right]^2 \rouge -\mu^2\
\noir \Phi^\dagger \Phi +\lambda \frac{v^4}{4}
\eeqn
Ideally one would like to explain how the minus sign for the
``{\em the mass of the doublet}" emerges. The Higgs interaction,
the masses and the mixing of the fermions and bosons are then
contained in

\beqn
{{\cal L}}_{0,1,1/2}=|D_\mu \Phi|^2 -V_{SSB} - \left( M_u^{ij}
\bar u_R^i \tilde{\Phi^\dagger} Q_L^j+ M_d^{ij} \bar d_R^i
\Phi^\dagger Q_L^j+ M_l^{ij} \bar l_R^i \Phi^\dagger L_L^j
 \;+\; h.c. \right)
\eeqn
Although  the masses of the $W/Z$ are derived from gauge
couplings, those of the fermions and the Higgs are Yukawa
couplings. The matrices in family space involve a large number of
parameters. Future probes of the flavour sector need to pin-down
these matrices and check whether the hierarchical mass and mixing
structures can be obtained in terms of very few parameters. For
instance, texture-zero matrices of the form\cite{fawzitexture}
\begin{equation}
\label{huhde} M_u=\lambda_t \left(\begin{array}{ccc} 0 & 0 & c \;
\epsilon^3 \; e^{i\phi}
\\ 0 & \lambda_c/\lambda_t & 0 \\
 c \; \epsilon^3  \; e^{-i\phi} &     0      &    1
\end{array}\right),\;\;
M_d=\lambda_b \left(\begin{array}{ccc} 0 & a \; \epsilon^3 & 0 \\
a \; \epsilon^3 & \epsilon^2 & b \; \epsilon^2 \\ 0 & b \;
\epsilon^2 & 1
\end{array}\right)
\end{equation}
with $a,b,c\sim 1$ successfully lead to $V_{us}=\sqrt{m_d/m_s}$.
Future precise measurements of $V_{ij}$ will drastically
discriminate between different ans\"atze. Recent evidence for
neutrino masses, may also be a sign of a new mechanism of mass
generation and hence of New Physics. The argument is theoretically
biased, even though very appealing. It is biased because although
it does not really explain why $\lambda_e/\lambda_t \sim 10^{-6}!$
it seeks to find a mechanism leading to a similar hierarchy factor
within a generation. Since a right-handed neutrino has no
electroweak quantum number and since neutrinos can get a Majorana
mass, neutrino masses may be induced to be small through the
see-saw mechanism. The latter gives the neutrino a tiny mass
because the right-handed neutrinos are extremely heavy, with a
mass scale related to the unification scale or some intermediate
scale, and therefore New Physics: $m_\nu=m_D^2/m_{\nu_R}$. With
$m_D$ the Dirac mass as generated for the quarks and charged
leptons and hence with a value similar to those, while $m_{\nu_R}$
is the extremely heavy right-handed neutrino mass.

While at it, note that the Higgs potential poses another
irritation. Potentially it contains a huge vacuum energy density
through the constant term in $V_{SSB}$. This can be written as
$V_{H,{\rm cosmos}}=\rho^H_{\rm vac}=M_H^2 v^2/8=\Lambda_H/8\pi
G_{\rm Newton}$. Even though recent results suggest $\Lambda \neq
0$, it rests that $\rho_{\rm vac} < 10^{-46}$GeV$^4$, while for
$M_h>100$GeV  $\rho_{\rm vac}^H> 10^{8}$GeV$^4$!! This problem is
still there in broken SUSY.\\
Talking about the cosmos, and assuming that our cosmological
models and other hypotheses are correct, then the matter content
of the \sm is not enough. There seems to be a large amount of dark
matter in the universe which calls for New Physics and new
particles.

\section{Lessons from the past:Why there should be New Physics}
\begin{figure*}[htbp]
\begin{center}
\mbox{\includegraphics[width=5cm,height=7cm]{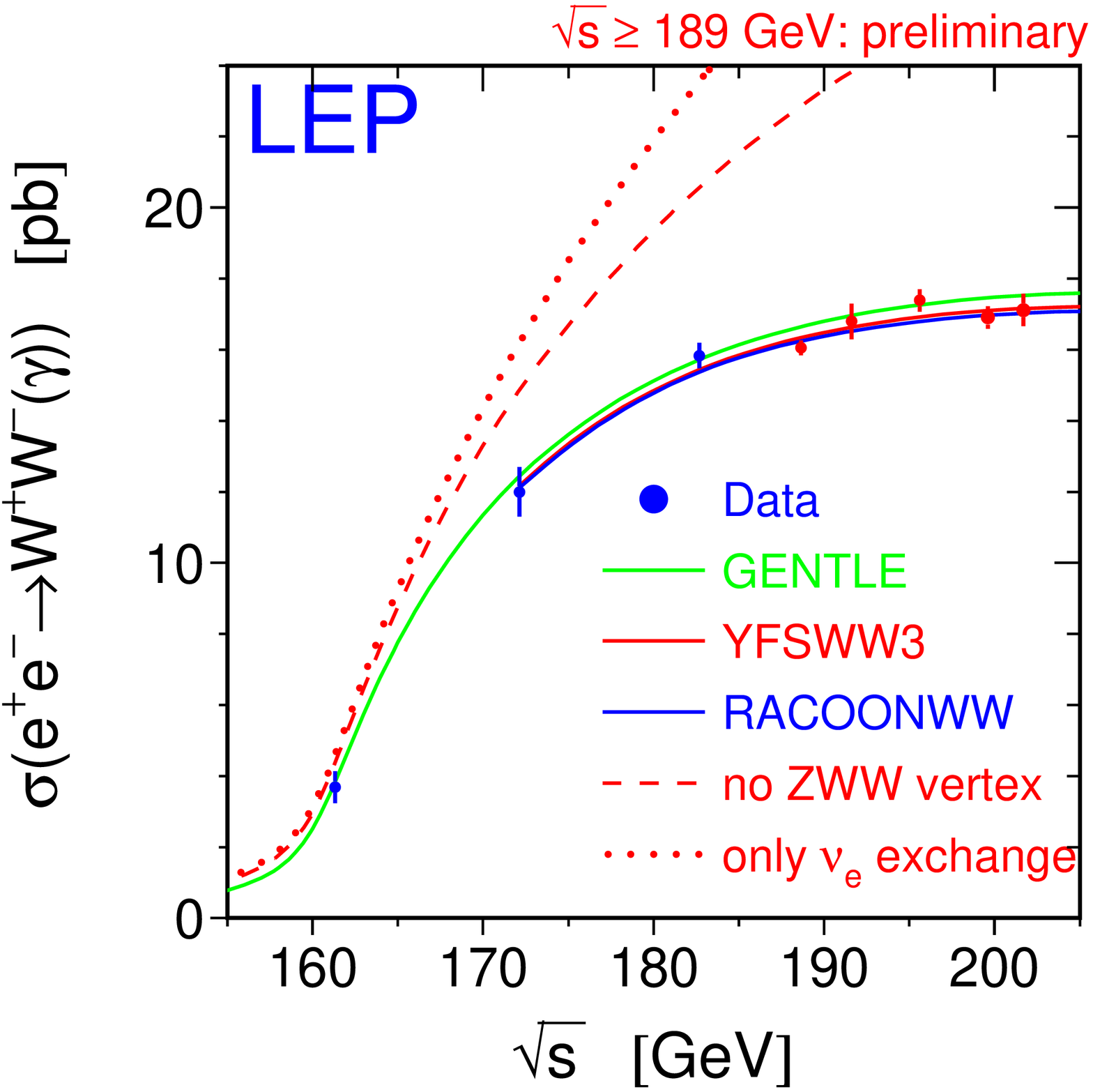}
 \mbox{\hspace*{-1cm}
\mbox{\includegraphics[width=12cm,height=8cm]{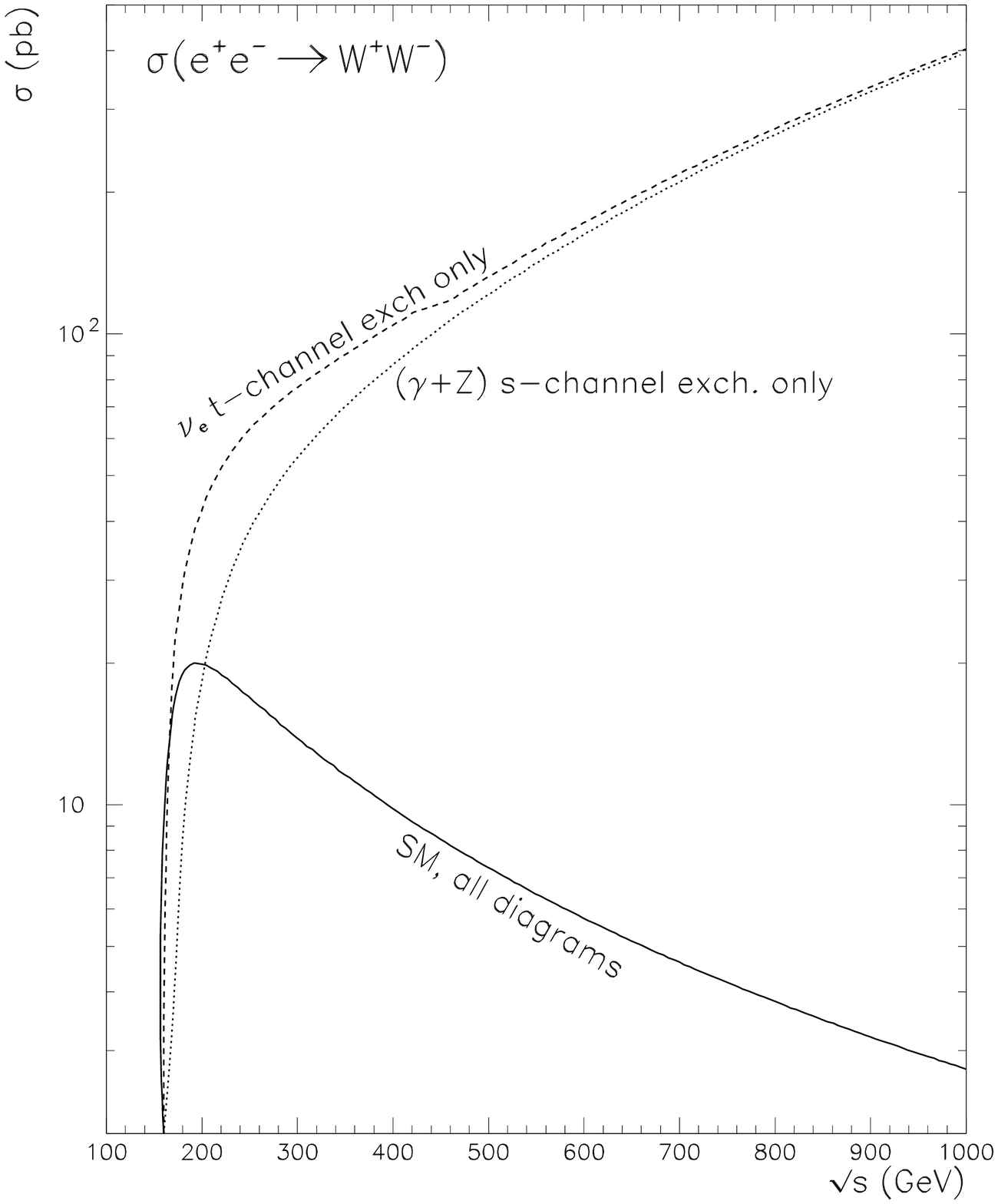}}
\raisebox{2.7cm}{ \mbox{ \hspace*{-8.5cm}
\includegraphics[width=6cm,height=2cm]{figeewwsm.eps}}}}}
\caption{\label{eetoww}{\em The small insert shows the latest data
of the $W^+W^-$ cross section at LEP2\cite{lepwwg}. The main
figure shows the behaviour of the same cross section at much
higher energies and the contribution of each channel\/.}}
\end{center}
\end{figure*}
Unitarity is a strong argument which is not based on some
theoretical prejudice or bias. This is just a statement that one
can not have a probability larger than one! This {\it generally}
translates into the fact that cross sections do not grow
indefinitely. A prime example is given by \eewwt as measured at
LEP2, Fig.~\ref{eetoww}. One sees that if the $WWV$ coupling and
the $e\nu_eW$ couplings are not equal, the cross section can get
catastrophically large. For this not to happen, either one has to
enforce the gauge symmetry at all energies or one has to invoke
New Physics to take place in case the $WWV$ coupling deviates from
its \sm value. Therefore in a sense the trend observed at LEP2 was
foreseen. This has happened time and again in the history of
physics. Fermi had described beta-decay  as a point-like
interaction. This had to break down. The reaction $\bar{\nu}_\mu
\mu \ra \bar{\nu}_e e$ calculated on the basis of the contact
interaction grows as $G_F^2 E^2$, $E$ is the c.m energy. Unitarity
would be violated at energies of the order $E > G_F^{-1/2} \sim
300$GeV. However physics that restored this behaviour, namely the
exchange of a $W$ boson, appeared much earlier: $80$GeV, while
precision measurements hinted at a departure from a point-like
structure even earlier. Next we could consider  $\nu_\mu
\bar{\nu}_\mu \ra W^+ W^-$ but with only the t-channel $\mu$
exchange, no $Z$ exchange. Unitarity in the $J=1$ channel would
have told us that unitarity would break at an energy $E
> \sqrt{3\pi/G_F} \sim 1TeV$. This is much higher than the LEP2
energy where the $WWZ$ coupling has been measured at better than a
few percents. In the situation we are in today we can consider
$W^+ W^- \ra W^+ W^-$
\begin{center}
\input{wwww}
\end{center}
and  take all the couplings to be gauge couplings. Adding all the
diagrams the helicity amplitude when all $W$'s are longitudinal
takes the form
\beqn
\label{ampliwwwwnoh}
{{\cal M}}_{LLLL} \sim \sqrt{2} G_F u
\eeqn
$u$ is one of the Madelstam variables. This shows that the cross
section will grow with energy. A partial wave analysis for the
$J=0$ channel shows that New Physics ought to be manifest at
$\sqrt{s_{WW}} \geq 1.2$TeV. This kind of energies require
post-LHC $pp$ machines and \epemt (or even $\mu^+ \mu^-$)
facilities operating in the range of some 3TeV or so and with
sufficient luminosity. This is not to say that a machine such as
TESLA will not be sensitive to a strongly interacting regime of
the weak interaction. In fact as we will see some useful
constraints can be set for some scenarios. But this is not always
guaranteed.\\
\noindent In the \sm, the picture can be improved by the inclusion
of the Higgs

\begin{center}
\input{wwwwH}
\end{center}
\vspace*{1cm}

turning the amplitude, Eq.~\ref{ampliwwwwnoh}, into
\beqn
{{\cal M}}_{LLLL} \sim -\sqrt{2} G_F M_H^2 \left(
\frac{s}{s-M_H^2} + \frac{t}{t-M_H^2}\right)
\eeqn
but again at asymptotic energies the amplitude grows with the mass
of the Higgs. This then puts a limit on the Higgs mass. Partial
wave analysis requires the {\em perturbative limit}
\beqn
M_H \leq \frac{4 \pi \sqrt{2}}{3 G_F} \sim 700GeV
\eeqn
within reach of the LHC.

\section{Lessons from the present and near future}
\begin{figure*}[htbp]
\begin{center}
\mbox{
\includegraphics[width=8cm,height=8cm]{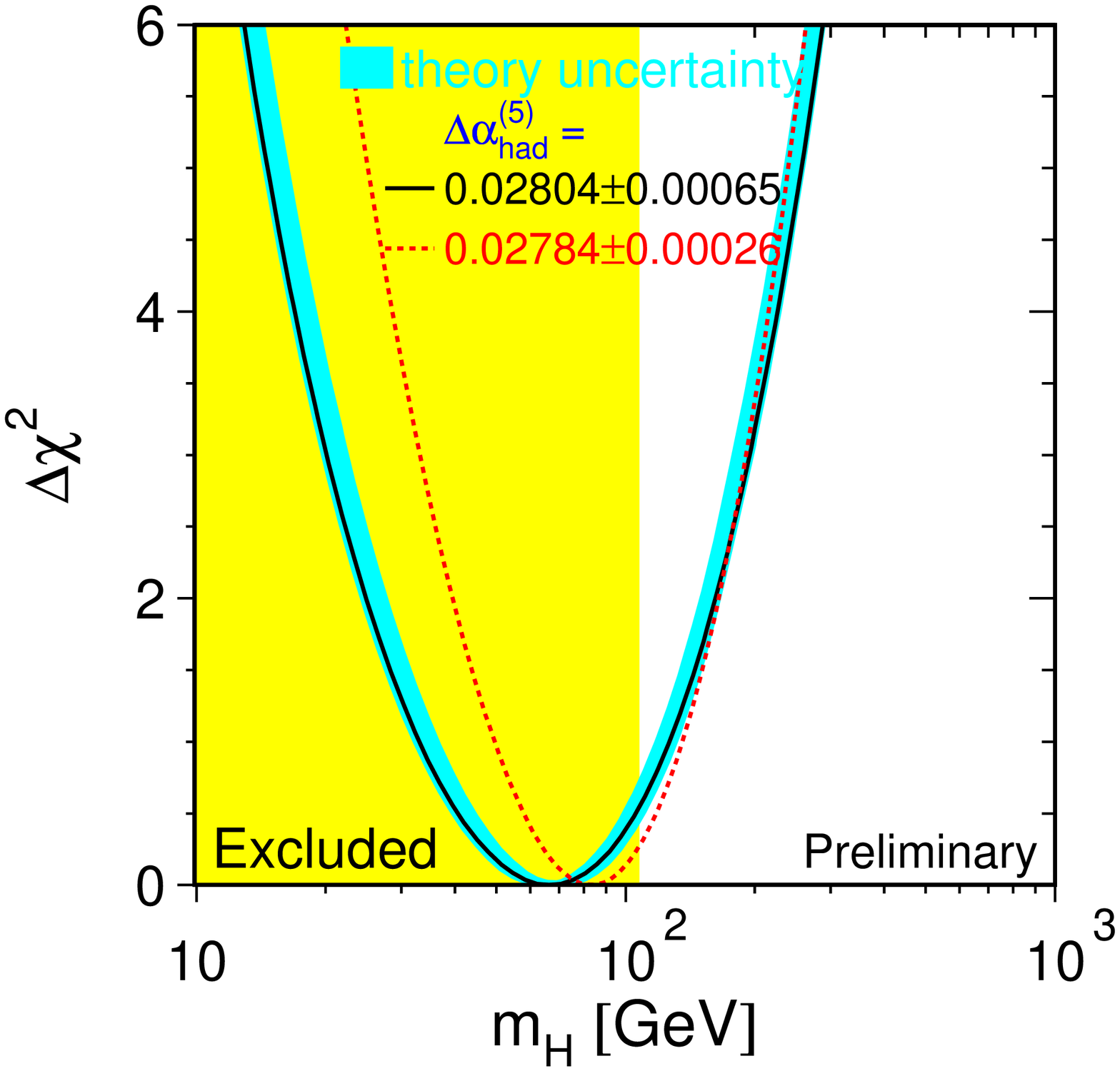}
\includegraphics[width=8cm,height=8cm]{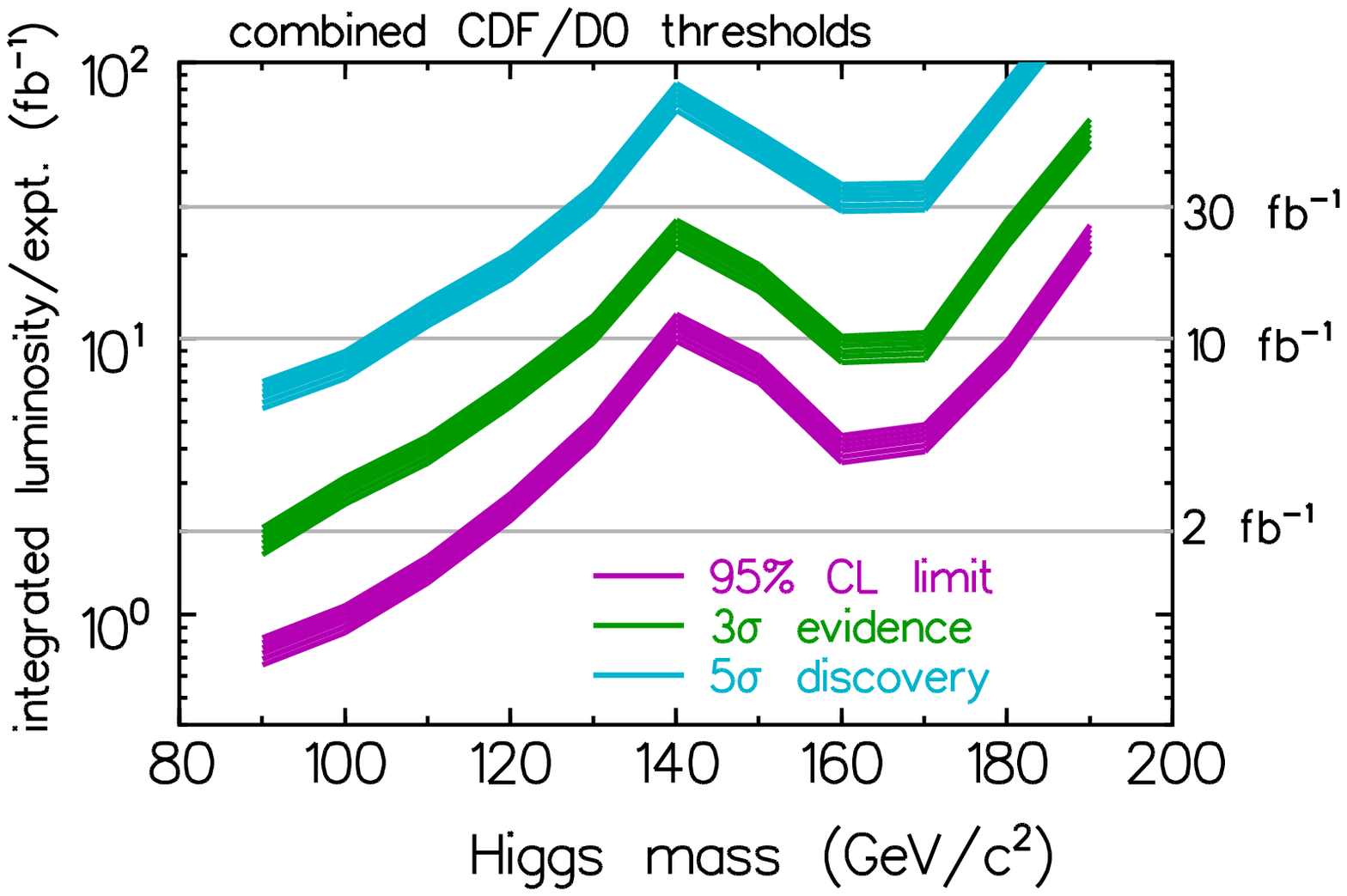}}
\caption{\label{mhlimit}{\em a) Latest limit on the \sm Higgs mass
from the precision measurements. The (yellow) shaded area gives
the direct limit\cite{lepwwg}. b) Expected discovery/exclusion
mass limit on the Higgs mass at the
Tevatron\cite{TevatronHiggsWG2000}\/.}}
\end{center}
\end{figure*}
Most people will ask why bother with this since LEP data indicates
the presence of a Higgs and constrains its mass to be less than
$170$GeV at $95\%$, Fig.~\ref{mhlimit}. A slightly higher limit is
derived if one uses improved values for $\alpha_{em}(M_Z)$.
Moreover many see the excess at the end of LEP2 as a tantalising
hint of a Higgs signal at $M_h\simeq115$GeV, that could be
confirmed by the Tevatron.

\begin{figure*}[htbp]
\begin{center}
 \rotatebox{90}{\mbox{
\includegraphics[width=6cm,height=10cm]{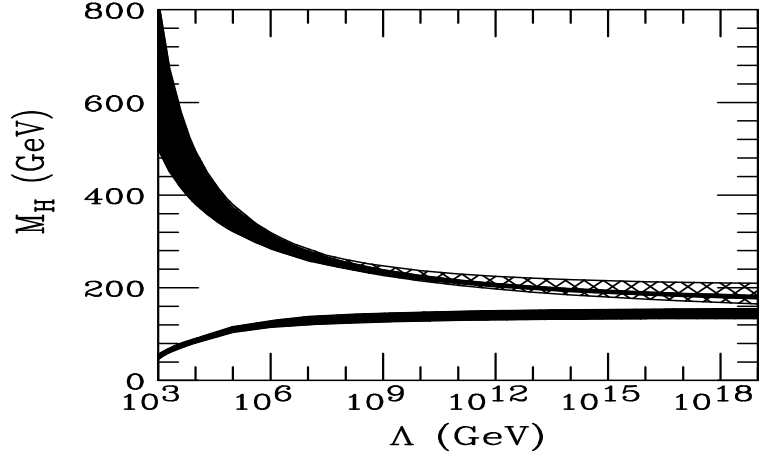}
}}
\end{center}
\vspace*{-1cm} \caption{\label{stability}{\em Requiring the Higgs
couplings to remain perturbative up to a certain scale $\Lambda$
gives the upper curve while imposing the vacuum to remain stable
leads to the other curve, from\cite{Riesselmann}\/.}}
\end{figure*}

Other arguments are often presented to back up the idea of a light
Higgs. Requiring the Higgs Yukawa coupling, and hence its mass, to
be perturbative up to a certain  scale $\Lambda$, where New
Physics should show up, and that the vacuum be stable give the
following constraint\cite{Riesselmann}: {\it \rouge If} \noir one
requires that the theory be perturbative up to the unification
scale and the vacuum be stable, then the Higgs mass is tightly
constrained around $180$GeV or so, see Fig.~\ref{stability}. A
Higgs mass as suggested by the excess seen at LEP2 would not be
compatible with a stable vacuum if $\Lambda > 10^{6}$GeV. Such a
light Higgs may then be a supersymmetric Higgs since the
contribution of the SUSY spectrum will make the vacuum stable at
scales compatible with the unification scale\cite{EllisDRoss}.
Note on the other hand that a heavy Higgs ($M_H>600$GeV) means
Physics at the TeV scale. So again either a light Higgs which when
combined with gauge coupling unification hints at new particles
around the TeV scale or a heavy Higgs but again with some
manifestation of New Physics at the TeV scale. In the first case
one {\underline {may}} be fortunate to discover, beside the light
Higgs, other new particles at the upcoming colliders, while in the
second case to reach the {\em TeV scale} {\underline {might}}
require a post-LHC hadron machine and/or a few TeV lepton
collider.

\begin{figure*}[htbp]
\begin{center}
\includegraphics[width=12cm,height=8cm]{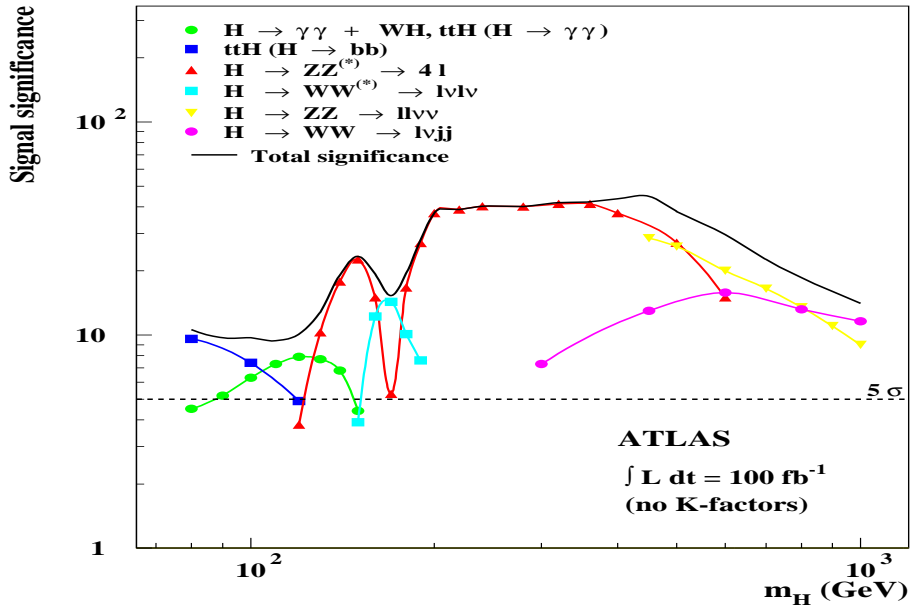}
\caption{\label{atlassmhiggs}{\em Statistical significance of a
\sm Higgs in ATLAS as a function of $m_h$. The various channels
are shown, from \cite{ATLAS_TDR}\/.}}
\end{center}
\end{figure*}

Even if the Tevatron fails to see a Higgs, the LHC should have no
problem discovering a Standard Model Higgs all the way from the
{\em direct} LEP2 limit to 1TeV, see Fig.~\ref{atlassmhiggs}. Note
however that the significance is lowest for $M_h \sim 115GeV$.
Moreover if the Higgs (especially a light one) is slightly non
standard, discovery may not be guaranteed. For instance invisible
decays of a light Higgs can easily bring the significance below
$5$. While at it, let us stress that within the MSSM, an invisible
Higgs invariably entails a rich production of neutralinos and
charginos because these would be quite
light\cite{nous_hinvisible_lhc}. One could then make precision
measurements on the masses of these to then explain why the Higgs
has been missed. Recent analysis suggest that even at the
LHC\cite{Zeppenfeld_invisibleH_lhc} (and the Tevatron if it is
produced\cite{Tevatron-invisible-lhc}) it may be possible to track
down an invisible Higgs. So wait... Of course an invisible Higgs
at the linear collider is no problem.

\section{Three solutions to the hierarchy problem and the three
main paths to the New Physics}

The previous discussions should be convincing enough as to why New
Physics should show up. Traditionally the main motivation for New
Physics is related to the hierarchy problem and again stems from
yet another frustration with the Higgs. Let us go once again
through the spin content of the \sm and the associated masses of
the particles. Why should the mass of the photon remain exactly
zero or why is it natural that those of the $W/Z$ are ridiculously
tiny compared to the what we think is the fundamental scale,
$\Lambda_P\sim 10^{19}$GeV? This is because these particles are
gauge bosons whose mass is protected by {\em local gauge
symmetry}. For the spin-$1/2$, a massless electron is protected by
a chiral symmetry. Though this is only a global symmetry it is
still powerful enough to keep the correction to the electron mass
tiny. Starting with a bare mass $m_e^0$, at one loop one gets
\beqn
m_e=m_e^0 \left( 1+\frac{3}{2} \frac{\alpha}{\pi}
\log(\Lambda^2/m_e^2) \right)
\eeqn
The correction is only logarithmic, taking the cut-off $\Lambda$
to be the Planck scale gives a $30\%$ correction. Doing the same
exercise for the scalar Higgs the correction for $M_h^2$ is
quadratic in $\Lambda$. If this is the Planck scale, this calls
for extraordinary fine-tuned adjustment close to 30digits! Clearly
this is unnatural and is related to the fact that there is no
symmetry that protects a scalar mass. Hence the motivation for the
New Physics based on the disparate scale of electroweak symmetry
breaking, $v \sim 246$GeV and $\Lambda_{\rm Planck}=\Lambda_P\sim
10^{19}$GeV.

\LARGE \rouge
\beqn
M_H\;,v \;\;\; \ll \;\;\; \Lambda_{\rm Planck} \nonumber
\eeqn

\normalsize \noir Notwithstanding that this is a numerical
accident and also that $\Lambda_{\rm Planck}$ relates to a force
which is not set on as firm foundation as the gauge theories of
the \smn, there are basically three main routes to solving the
hierarchy problem. Each one tackles one side or one part of the
above inequality.

\rouge i) $M_H$ just not there! \noir These models require no
Higgs or rather no fundamental elementary scalar. The Higgs may be
a composite particle that can be heavy. As we have seen earlier
these models would suggest that the $W$ interaction can become
strongly interacting at the TeV.\\

\rouge ii) $\Lambda_{P}$ is not the fundamental scale. \noir The
fundamental scale is much lower and can be at the TeV scale. This
is the very recent extra-dimension solution. It does not require a
host of new particles. Although this brings the fundamental scale
down to $v$, little has been convincingly done to dynamically
induce $v$ from the fundamental scale. If this is not done, the
fact that the two scales refer to quite different mechanisms,
leaves a puzzle as to
why they are so close to each other. For some recent attempts,
see \cite{KKtops}\\

\rouge iii) $\ll$ is quite natural. \noir This is the
supersymmetric solution and seems to me more satisfying than the
previous solution since it endows the scalar with a symmetry that
protects its mass, by associating the scalars to fermions that
have a protective mechanism. By doubling the known spectrum, the
phenomenology is very rich.

\subsection{The Higgsless models}
First of all, it is worth stressing that one can have a perfectly
gauge invariant theory without a Higgs. One only needs to make use
of the Goldstones. One should also take into account the fact that
the $\rho$ parameter being to a good approximation unity suggests
a custodial symmetry. Then use (for notations and a mini review
see \cite{Morioka})
\beqn
\Sigma=exp(\frac{i \omega^i \tau^i}{v}) \;\;\; {{\cal D}}_{\mu}
\Sigma=\partial_\mu \Sigma + \frac{i}{2} \left( g \W_{\mu} \Sigma
- g'B_\mu \Sigma \tau_3 \right)
\eeqn
The $W,Z$ masses are simply
\beqn
\cl_M=\frac{v^2}{4} \tr(\cd^\mu \Sigma^\dagger \cd_\mu \Sigma)
\eeqn
which is the lowest order operator one can write. \\
The severe problem this model faces, is how to accommodate the LEP
data that calls for a light Higgs. The answer is that those fits
are only valid within the \smn. Fitting the data with a larger
Higgs mass calls for New Physics contributions to the LEP
observables. In a more general context one has to consider the
$S,T,U$\cite{PeskinSTU} (or
$\varepsilon_{1,2,3}$\cite{AltarelliBarbieri_epsilon}) variables
besides the Higgs mass. These can be approximated as
\beqn
\label{STUvariables}
 \varepsilon_1&=&\Delta\rho=\alpha T=\frac{3G_\mu M_Z^2}{8
\pi^2 \sqrt{2}} \left( \frac{m_t^2}{M_Z^2}-2 s_Z^2 \ln(M_h/M_Z)
\right)
+\varepsilon_1^{\rm NP} \nonumber \\
\varepsilon_2&=&-\frac{\alpha}{4 s_Z^2} U=-\frac{G_\mu M_W^2}{2
\pi^2 \sqrt{2}} \left( \ln(\frac{m_t}{M_Z}) \right)
+\varepsilon_2^{\rm NP} \nonumber \\
\varepsilon_3&=&\frac{\alpha}{4 s_Z^2} S=\frac{G_\mu M_W^2}{12
\pi^2 \sqrt{2}} \left( -\ln(\frac{m_t^2}{M_Z^2})+
\ln(M_h/M_Z)\right) +\varepsilon_3^{\rm NP}
\eeqn
In models where the Higgs is absent, $M_h$ in the above should be
interpreted as a cut-off at the TeV scale ($1$TeV to $4\pi v\sim
3$TeV). The contribution which is most sensitive to the Higgs is
$\varepsilon_3$. In $\varepsilon_1$ the Higgs dependence is
somehow subleading compared to the quadratic top mass dependence.
The $S$ variable has been a killer of naive technicolour.
Precision measurements interpreted with a large Higgs mass need
$S_{\rm new}<0$ to counterbalance the large Higgs mass/cut-off.
Recent fits, allowing $T$ as free parameter, and taking the New
Physics at $3$TeV give\cite{BaggerSwartzSTU} $-.3 <S_{\rm
new}<-.1$. Technicolour based on a naive rescaling of QCD, leads
to $S>0$. In an effective Lagrangian approach including
next-to-leading operators to those contributing to the mass, one
should expect the following operator, beside an operator that
accounts for a slight breaking of the custodial symmetry and hence
to $T$,
\beqn
{{\cal L}}_{10}&=&g g' \frac{L_{10}}{16 \pi^2} \tr ( \B^{\mu \nu}
\Sigma^{\dagger} \W^{\mu \nu}  \Sigma ) \longrightarrow
L_{10}=-\pi S_{\rm New}
\eeqn
Limit from LEP/SLC thus suggest \rouge $L_{10} \sim {{\cal
O}}(.1)$\noir . One can therefore fit the LEP data with a very
heavy Higgs by including a ``judicious" amount of $S$ and $T$.
\noir Some have argued that one should also consider a larger set
of operators that involve the various fermion
fields\cite{BarbieriStrumia_effOper}, and not just those bosonic
that contribute only to the two variables $S$ and $T$. The
fermionic operators give extremely constraining bounds and thus if
{\it all} operators are of the same order, which could be argued
is what is ``natural", then one can not so easily fit a large
Higgs mass. However it may well be that the correct effective
theory leads to negligible fermionic operators. It is not a very
attractive possibility but one that can not be totally dismissed.
Nonetheless, considering solely the bosonic operators, one can
write others beside $L_{10}$. Although they do not contribute
directly to the present precision measurements they do contribute
to the tri-linear and quadri-linear vector boson couplings. For
instance, ${{\cal L}}_{9L}=-i g \frac{L_{9L}}{16 \pi^2} \tr (
\W^{\mu \nu}\cd_{\mu} \Sigma \cd_{\nu} \Sigma^{\dagger} )$ and
$\cl_{1}=\frac{L_1}{16 \pi^2} \left( \tr (D^\mu \Sigma^\dagger
D_\mu \Sigma) \right)^2$ to cite only two (for more see
\cite{Morioka}). Now these operators need to be probed at higher
energies. In order that one learns more than what we have with the
LEP data, these operators should be constrained better than
$L_{10}$, {\it i.e.}, the $L_i$ should be measured better than
$.1$, ideally one should aim at the $10^{-2}$ level. This is hard
since already $L_{9L}\sim .1$ implies measuring the $\Delta
\kappa_\gamma$ in the $WW\gamma$ vertex at $\Delta \kappa_\gamma
\sim 1.3 \;10^{-4}$

\begin{figure*}[htbp]
\begin{center}
{\mbox{
\includegraphics[width=10cm,height=10cm]{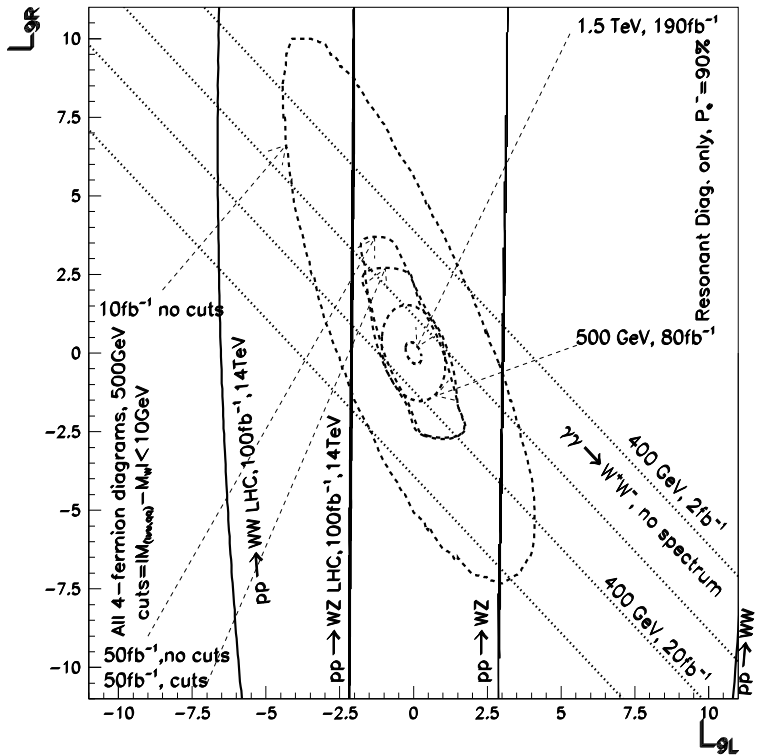}
}}
\end{center}
\caption{\label{l9limit}{\em Comparison of limits on the chiral
Lagrangian parameters $L_9$ at the future colliders\/,
from\cite{Morioka}.}}
\end{figure*}

The figure shows how the tri-linear couplings would be measured at
the colliders. \epemt even at relatively modest energies does a
good job, but one is still somehow below the sensitivity one has
reached on $L_{10}$. Recent simulations for TESLA ($500$GeV) with
improved luminosity ($500$fb) give $L_{9} \sim .2$ while $pp \ra
WZ$ at the LHC give $L_{9L} \sim 1$\cite{KilianWWWW}. $WW$
scattering will constrain the quartic couplings $L_{1,2}$. For
these one definitely needs an \epemt in excess of a TeV, the
precision on the couplings scales as $s^{-1} \sqrt{{{\cal
L}}^{-1}}$. A simulation for 1TeV \epemt machine with a luminosity
of $500$fb$^{-1}$ give a limit $L_1 \sim 1.5$ while
the LHC gives $\sim 5$\cite{KilianWWWW}. \\
Once again if one takes the $L_{10}$ limit set by the  present
precision measurements as a yardstick, one would again come to the
conclusion that the physics of a heavy Higgs needs a post-LHC and
a  post-LC500GeV. In fact if the latter can run as a
Giga-Z\cite{GigaZ} factory with polarised beams and with an
improved measurement of the $W$ mass and the top mass ($\Delta m_t
\sim 200$MeV is foreseen), then the $S,T,U$ variables will be
incredibly constraining dwarfing the importance of the
measurements of the other $L_i$'s from $W$ pair production. To be
fair and stress the importance of future measurements, one should
also add that it is also not excluded that $L_{10}$ be
``naturally" much smaller than the other $L_i$'s. Indeed the
couplings we have written invoke the custodial $SU(2)$ symmetry.
One could impose an additional global symmetry
$SU(2)_A$\cite{InamiL10} which would make $L_{10}$ vanish. For
instance in some disguises of technicolour this would occur if the
vector and axial vector mesons had the same mass. This is
implemented in the extended BESS model\cite{SuperBESS}, where
$L_{10}$ is zero at the leading order. The model enjoys then a
decoupling property, where the contributions to the $S,T,U$
variables become all of the same order and are subleading.
Nonetheless to be consistent with a heavy Higgs, means that these
contributions are not negligibly small and thus that the vector
bosons are not asymptotically heavy. They should appear as
resonances at the LHC or perhaps even at the LC.

\subsection{Extra-dimensions}
This scenario solves the hierarchy problem by suggesting that the
natural fundamental scale is not the Planck scale with $\Lambda_P
\sim 10^{19}$GeV but of the order a few TeV! This is based on the
reasoning that if all matter and interactions were indeed
occurring in a 4-dimensional world but that gravity were all
pervading in a larger dimensional space, gravity will then look
weaker to us while actually it is not\cite{ADD}. As for the
extra-dimensions one could think of them as being compactified
with a ``small" radius $R$. For distances $r$, with $r \gg R$ one
would have the usual (4-dim) Newtonian law
$$F \propto \frac{1}{\Lambda_P^2} \frac{m_1 m_2}{r^2}$$
($\Lambda_P^2=4 \pi G_N^{-1}$). However for $r \ll R$ one should
feel the $4+n$ law $$F \propto \frac{1}{\Lambda_{F}^{2+n}}
\frac{m_1 m_2}{r^{2+n}}.$$ By matching \rouge $ \Lambda_P^2=
R^n\Lambda_F^{2+n}$ \noir. In the original ADD variant\cite{ADD},
it was required to have the fundamental scale $\Lambda_F \sim v
\sim 1TeV \sim 10^{-17}cm$. This already excludes $n=1$ since we
extract $R \sim 10^{15}$cm, {\it i.e.} distances where the $1/r^2$
law has been extremely well tested. $n=2$ gives $R\sim mm$, it is
still not totally excluded. $n=5$ corresponds to $R\sim fm$. In
this scenario the most striking signature is the production of
gravitons, G, that could pop out of the detector into the extra
world and act as missing energy. The most promising reactions are
$\epem \ra \gamma G$ and $pp \ra g G$. Although gravitons still
interact with a strength $1/\Lambda_P$, there is a huge density of
them. The scaling law for the cross section can be explained
easily. Take the simple case of an extra dimension  with a simple
geometry where the extra-dimension is compactified to a circle
with radius R. The wave function can be factorised as
\beqn
\Psi_G (x_4,y) =\sum_k \Psi_k (x_4) e^{i k y/R}
\eeqn
From the point of view of the 4-dim world one has a tower of
Kaluza-Klein states of mass $m_{G,k} \sim k/R$ and thus one can
have for a total available energy $\sqrt{s}$ about $N=\sqrt{s}R$
states contribute to the cross section. For $n$ extra dimension
one has $N=(\sqrt{s}R)^n$, and therefore though the gravitational
interaction is of order $1/\Lambda_P$, the cross section is
$\sigma \sim N \times 1/\Lambda_P^2 \sim 1/\Lambda_F^2 \times
(s/\Lambda_F^2 \times)^{n/2}$. Limits one expects are given in the
table below\cite{PeksinExtradimLimit}.

\begin{center}
\begin{tabular}{|c|c|c|}
\hline
   & n=4 & n=6 \\
   & R(cm)/$\Lambda_F$(GeV) & R(cm){\bf /}$\Lambda_F$(GeV) \\
   \cline{2-3}
  {\bf LEP} & $1.9 \; 10^{-9}${\bf /} 730 & $6.8 \; 10^{-12}${\bf /} 530\\
  {\bf LHC} & $5.6 \; 10^{-11}${\bf /} 7500 & $2.7 \; 10^{-13}${\bf /} 6000 \\
  {\bf LC500} & $1.2 \; 10^{-11}${\bf /} 4500& $6.5 \; 10^{-13}${\bf /} 3100 \\ \hline
\end{tabular}
\end{center}

Gravitons can also contribute indirectly leading to contact
interactions but the limits are model dependent. What about the
Higgs in these models. Here the Higgs mass is a free parameter,
nonetheless though most studies concentrate on the $J=2$
component, it is conceivable that the $J=0$ mixes with the Higgs
on the brane and leads to a decay of the Higgs into
graviscalars\cite{Htograviscalars} and hence an invisible Higgs
signature. This contribution can be large for a light Higgs. Both
the LHC and the Tevatron could well miss the signal, but not the
LC. One can of course also imagine that not only the gravitons but
also the ordinary gauge bosons have siblings as Kaluza-Klein
towers. For that one can assume the gauge bosons to propagate in
more than 4-dim provided on takes for instance $R \sim 1 TeV \sim
10^{-17}$cm, $\Lambda_F\sim 10^4$TeV. The KK states of would be
excited $W'/Z'$  could be looked for at the colliders. Moreover
these states can mix with the ordinary vector bosons and
indirectly contribute to the $\varepsilon_{1,2,3}$,
Eq.~\ref{STUvariables} if not too heavy. In this case one could
fit the LEP data with a heavier Higgs
($M_h<500$GeV)\cite{WellsRizzo} but would expect to see the
additional KK gauge bosons at the LHC or LC500. It rests that,
though the extra-dim scenario is in its infancy,  it does not
guarantee that one will see its manifestation. It may well happen
that the Higgs (well at least that) is discovered at the LHC in
accordance with the LEP limit and that it will not pose us a
naturalness problem within this scheme...

\subsection{SUSY}
The SUSY solution is in my view a better solution to the hierarchy
puzzle since it solves the problem  through a symmetry which alas
 has to be broken. Let us start with the Higgs sector of the
theory. A SUSY version of the \sm requires, for anomaly
cancellation, two Higgs doublets ($H_{1,2}$). Proper symmetry
breaking then gives 5 physical Higgses: 2CP even $h,H$, one CP odd
$A$ and the charged Higgs $H^\pm$. But to do that SUSY must be
broken. Indeed, before SUSY is broken the general supersymmetric
potential is
\beq
V = |\mu|^2 \left(
  |H_1|^2 + |H_2|^2 \right) + \frac{g^2+g'^2}{8} \left( |H_1|^2 -
  |H_2|^2 \right)^2 + \frac{g^2}{2} |H_1^*H_2|^2 \geq 0\nonumber
\eeq

Note the appearance of the $\mu$ term which is a SUSY conserving
{\em free} parameter. But note also that the quartic couplings are
gauge coupling. So one must add (soft) SUSY breaking parameters in
such a way that one triggers electroweak symmetry breaking.
\bea V_H
&=& (m_{11}^2|+\mu|^2 ) |H_1|^2 + (m_{22}^2+|\mu|^2)  |H_2|^2 -
m_{12}^2 \epsilon_{ij} \left( H_1^i H_2^j + h.c. \right) \nonumber
\\ &+&
\frac{g^2+g'^2}{8} \left( |H_1|^2 - |H_2|^2 \right)^2 \;+\;
\frac{g^2}{2} |H_1^* H_2|^2
\eea

With {\em appropriate} soft-susy breaking terms electroweak
symmetry breaking can be achieved. As this introduction to the
Higgs potential already shows, understanding how supersymmetry is
broken will be a prime motivation if supersymmetry is discovered.
It will be crucial to reconstruct as much as possible the
soft-susy parameters and test whether they satisfy some specific
relations. A  host of soft-susy breaking masses and mixing are
also necessary in order to explain the splitting between the
masses of the fermions and the sfermions. In all generality these
masses and $A$ terms should be matrices in flavour space. However
because of the danger of potentially large FCNC one has to retort
to the assumption of some alignment, {\it i.e.} diagonal matrices
(the same basis as that of ordinary fermions). Most popular models
even assume a common scalar mass for all and a common tri-linear
$A$ term at some unification scale. For each gaugino one must also
attribute a soft susy breaking mass. It is also common to assume
that the three gaugino masses also unify at the unification scale
in the same way that within SUSY the three gauge couplings unify.
All these assumptions need to be verified because they are probing
theories and mechanisms at GUT scales or even string scales!
\begin{figure*}[htbp]
\begin{center}
\mbox{\includegraphics[width=10cm,height=8cm]{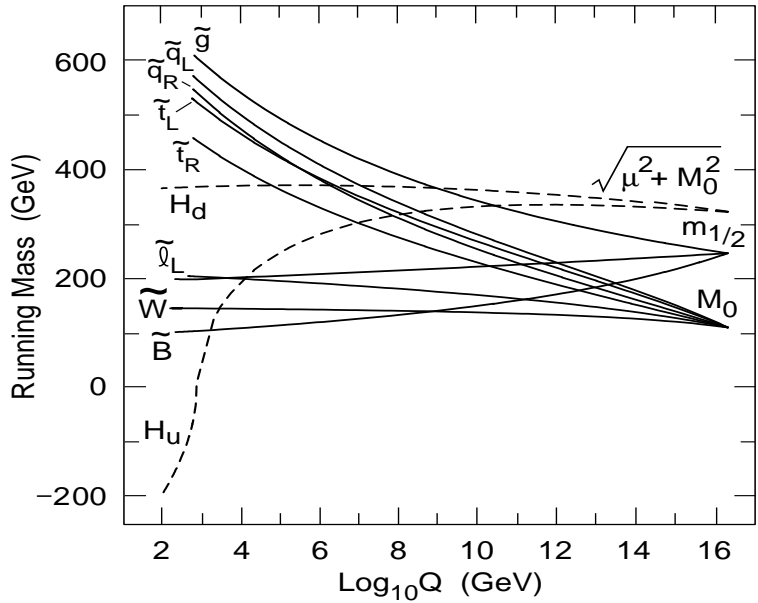}}
\caption{\label{sugraevolv}{\em Evolution of the the SUSY masses
in mSUGRA.From \cite{evolution_susymass}\/.}}
\end{center}
\end{figure*}
One popular model with a minimum number of parameters is the
mSUGRA\cite{mSUGRA} model where all scalar masses are equal at the
unification scale as are all the gaugino masses. The SUSY spectrum
it predicts at the weak scale is quite distinctive and is shown
above, Fig.~\ref{sugraevolv}. One extremely appealing feature is
that one can induce at the weak scale a ``negative mass" for $H_2
=H_u$! This is however not always guaranteed and even if possible
and it may be too fine-tuned (the $\mu$ problem).  Moreover it is
not always assured that some other charged particles do not end up
with an unwanted negative mass!. Still the reconstruction of the
fundamental SUSY parameters will be a fascinating subject if SUSY
is indeed discovered. Well at least one of the  Higgses should be
discovered.

\begin{figure*}[htbp]
\begin{center}
\mbox{
\includegraphics[width=8cm,height=8cm]{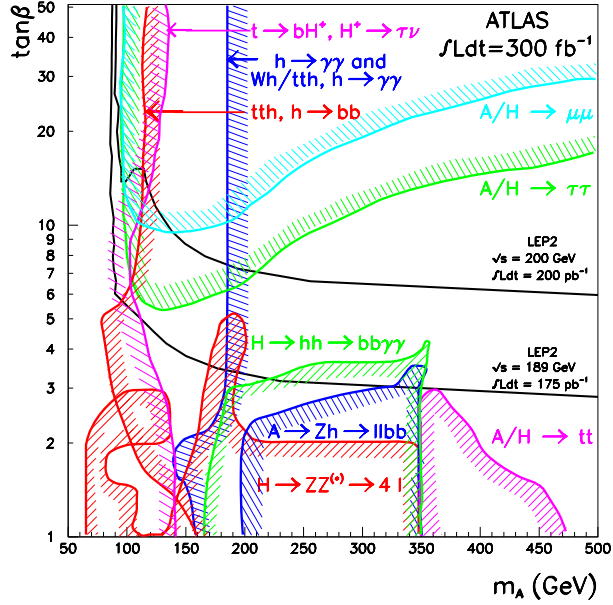}}
\caption{\label{AtlasSusyHiggs}{\em a Discovery potential of SUSY
Higgses at the LHC. All other sparticles are assumed heavy\/, \cite{ATLAS_TDR}.}}
\end{center}
\end{figure*}

At tree-level, and in the large $M_A$ limit, one has $m_h^2=M_Z^2
\cos^2 2 \beta$ which is excluded by LEP. The tree-level relation
receives a large radiative correction due to the large top Yukawa
coupling and the contribution of the stops, but one still has a
definite upper limit $m_h < 130$GeV\cite{SUSYHiggsmass}. In models
beyond the MSSM, this upper limit is relaxed. However insisting
that all Yukawa couplings remain perturbative up to the
Unification scale, then one finds that
$m_h<205$GeV\cite{mhsusy205}. At LC500 such a light Higgs will be
discovered within a day! Even if it decays invisibly. At the LHC,
since the main signature is in $\gamma \gamma$ it is not so easy
but all simulations show that one should not miss the Higgs and if
lucky might even see some of the others, see
Fig.~\ref{AtlasSusyHiggs}.

The above picture, Fig.~\ref{AtlasSusyHiggs}, is with the
assumption that the other SUSY particles are too heavy to have an
impact on the decay patterns and the production mechanisms.
However if SUSY is correct it is very unlikely that the other
particles are too heavy. One can for instance have light stops
with large mixing that can reduce drastically the $gg \ra h$
production\cite{RggKane,Kileng_mixing,AbdelStop_Hgg_Loops,nous_Rggstophiggs_lhc}.
But the same scenario not only enhances the associated production
signal but also guarantees discovery of $\sto$ and may even
generate higgses through $\stt \ra \sto h
$\cite{nous_Rggstophiggs_lhc}. Another potential danger is the
possibility of $h \ra \neuto \neuto$, that decays into invisible
LSP's\cite{nous_hinvisible_lhc}. But again this means that there
will be a nice study of the chargino neutralino system at the LHC
and perhaps even at the Tevatron. Also in mSUGRA, it is very
possible to produce the Higgs in the decay chain $\neutt \ra
\neuto h$\cite{ATLAS_SUSYtoh}. So light sparticles are often a
blessing for the Higgs. The picture may be more complicated for
decays of the heavier Higgses.

\begin{figure*}[htbp]
\begin{center}
\mbox{
\includegraphics[width=8cm,height=8cm]{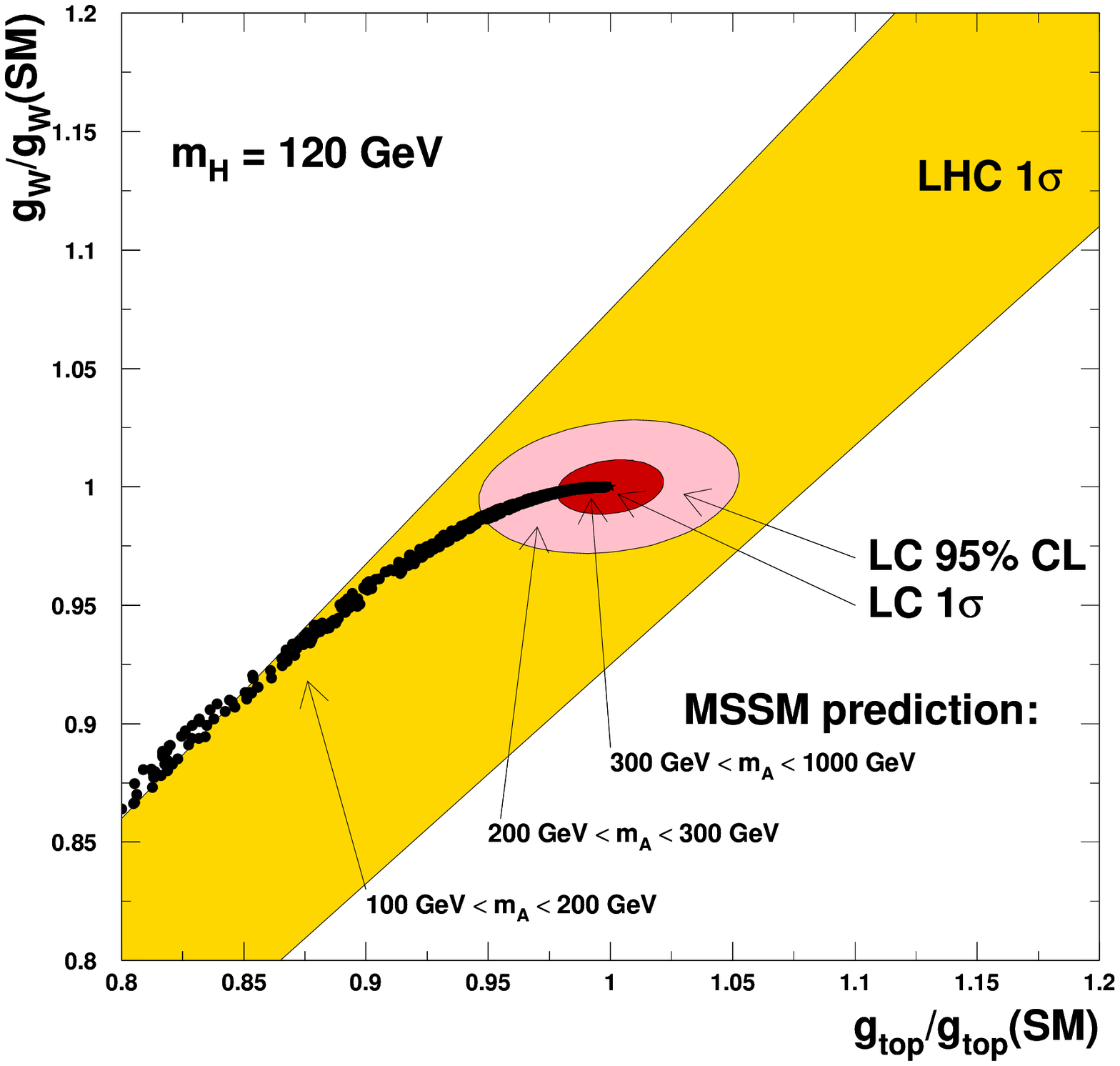}
\includegraphics[width=8cm,height=8cm]{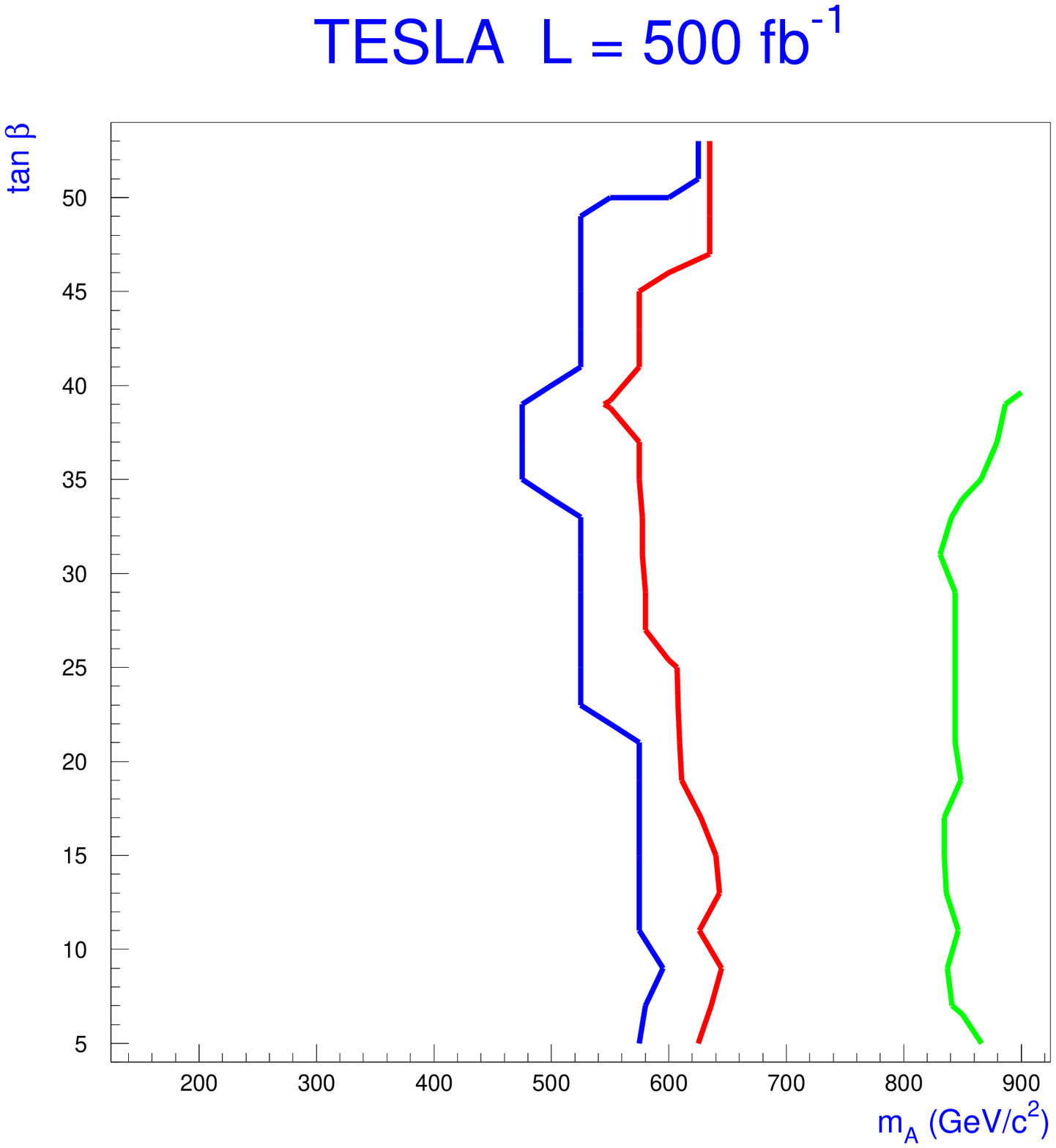}}
\caption{\label{LChiggs}{\em a) Precision with which the Higgs
couplings $h t\bar t$ and $h WW$ couplings, normalised to those of
the \sm, can be measured at TESLA and LHC and how these parameters
vary with $M_A$ in SUSY. b) Distinguishing a SUSY $h$ apart from a
\sm on the basis of a combination of Higgs branching ratios at
TESLA with $1000$fb$^{-1}$. The sensitivity in $M_A/\tgb$ is shown
at $68\%$, $90\%$ and $95\%$ level. From
\cite{BattagliaDesch}\/.}}
\end{center}
\end{figure*}
Although the latter may not be directly accessible at the
LC500\footnote{The  \gamgamt mode can extend the discovery
potential of the heavier neutral Higgses. For a review on the
$\gamma \gamma$ mode see \cite{Parisgg}. For a recent study on the
detection of the heavier neutral Higgses see
\cite{HeavierHinLaser} } the bounty of $h$ in the very clean
\epemt environment allows first class precision measurements on
the Higgs properties. The spin and parity of the Higgs can be
measured. Extraction of the {\em light} Higgs couplings to
fermions (and $W$'s) at the $1\%$ level (compared to $10-15\%$ at
the LHC) can distinguish between a \sm Higgs and a SUSY Higgs and
may either set a limit on $M_A$ ($M_A
>600$GeV) or constrain its mass if not too heavy$M_A<500$GeV, practically
independently of \tgbt, see Fig.~\ref{LChiggs}.

\begin{figure*}[htbp]
\begin{center}
\mbox{
\includegraphics[width=8cm,height=8cm]{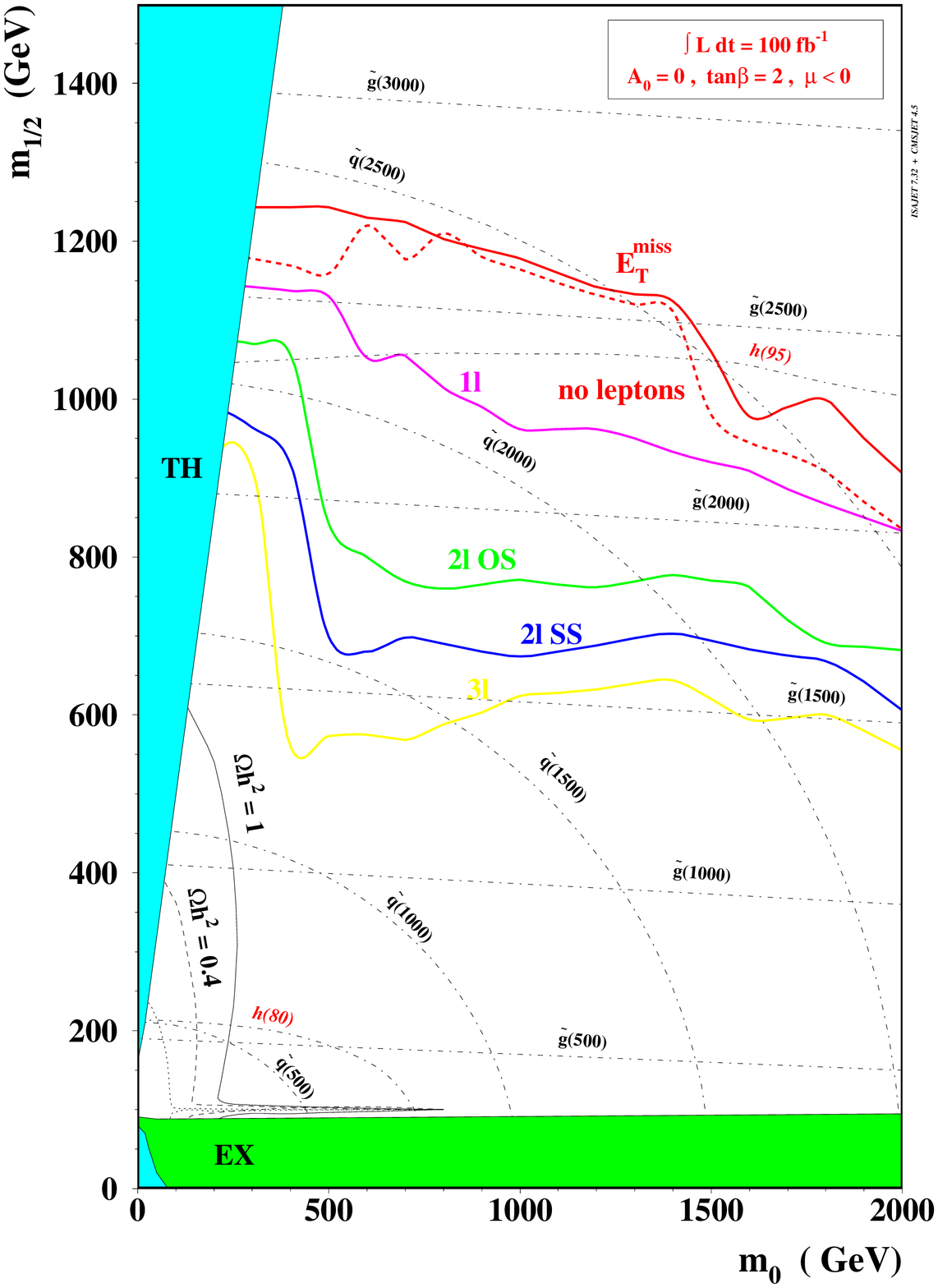}
\includegraphics[width=8cm,height=8cm]{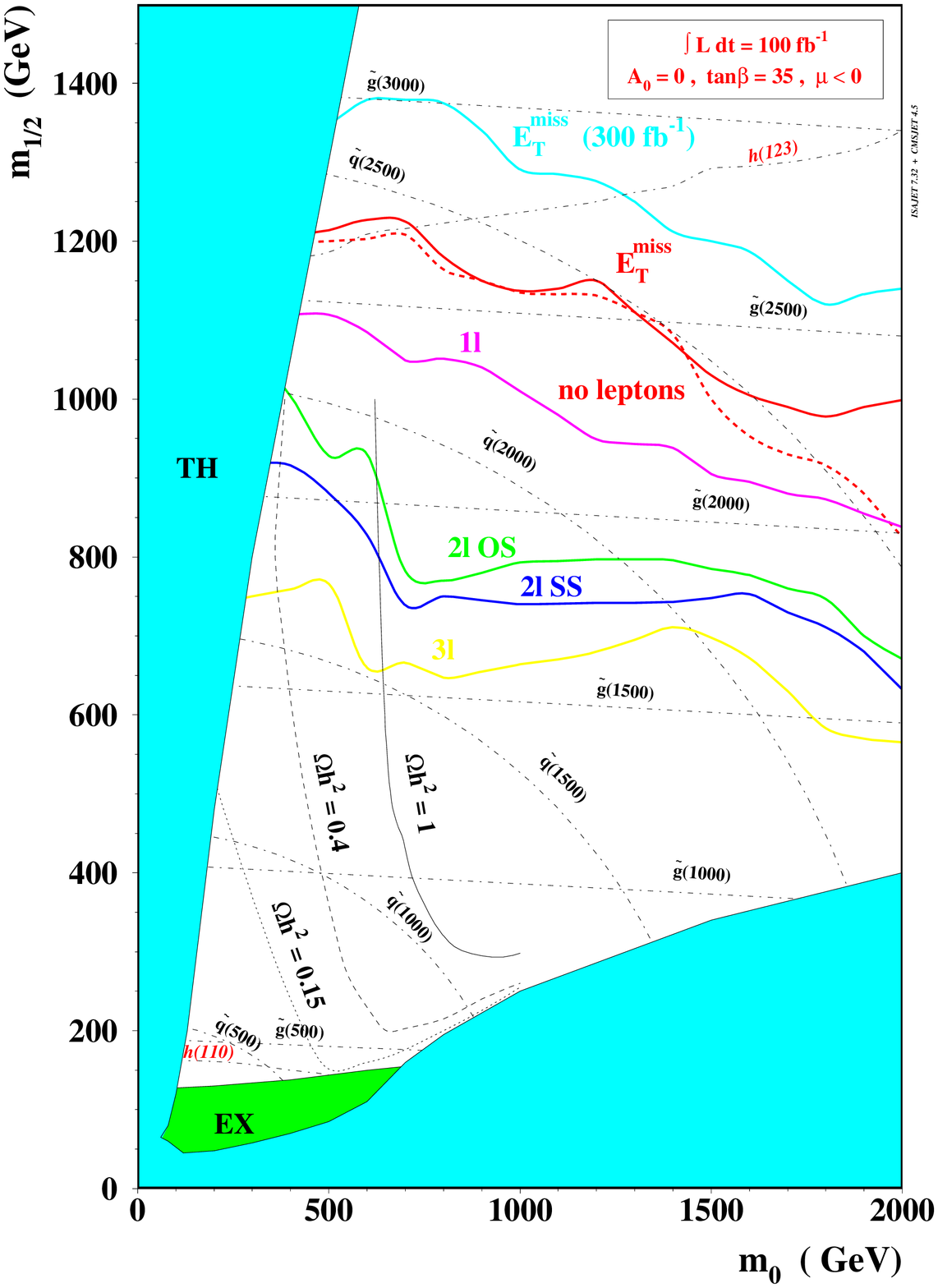}}
\caption{\label{cmsusy}{\em Discovery potential of squarks and
gluinos in CMS in the $m_0,M_{1/2}$ plane of mSUGRA for two values
of \tgbt . The $5\sigma$ contours corresponding to different final
state signatures are shown. Isocurves for the masses are also
shown. Note that constraints on the relic density, and the Higgs,
are also shown. These would make these models not viable, however
these can be circumvented in slightly more general that give the
same  LHC reach for gluinos and squarks, From
\cite{AbdullinCharles}\/.}}
\end{center}
\end{figure*}
\begin{figure*}[htbp]
\begin{center}
\includegraphics[width=8cm,height=2cm]{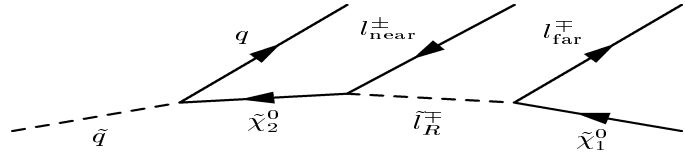}
\caption{\label{decaychain}{\em A typical decay chain for the
squark \/, from \cite{decaychainLHC}}.}
\end{center}
\end{figure*}

Even if for $M_A>600$GeV one one would see only the lightest Higgs
at both the LHC and LC500, squarks and gluinos with masses as high
as $2-3$TeV should be accessible at the LHC in a number of
signatures, as Figure~\ref{cmsusy} shows. Apart from some quixotic
scenarios, failing to see squarks and gluinos means that they have
masses in excess of $3$TeV. This would cast some very serious
doubt about SUSY as  providing a neat answer to the hierarchy
problem. Therefore if SUSY is correct and with such a large number
of squarks produced one needs to precisely reconstruct the SUSY
parameters. Until recently this has been done in the LHC
simulations by assuming a specific model, most often mSUGRA.
Recently many of the model assumptions have been dropped, however
one still needs to rely on a specific decay chain to be able to do
anything useful\cite{SUSYLHC}. Take for instance the following
chain, Fig~\ref{decaychain}.

By measuring the different combinations of invariant mass
distributions and studying their end-points it is possible to
measure the SUSY masses with a precision close to a few per-cent.
In some cases only mass differences can be extracted, but even
these can be very useful in distinguishing between models.

Nonetheless the conclusions would still be biased. For instance,
one can not, in a model-independent way,  claim that the slepton
is the right-handed slepton, and it may be even not easy to
ascertain its spin. Even more difficult, how do we know it is a
supersymmetric process? A confirmation would be to measure the
coupling $g_{\ser e \neuto}$.

\begin{figure*}[htbp]
\begin{center}
\mbox{
\includegraphics[width=8cm,height=8cm]{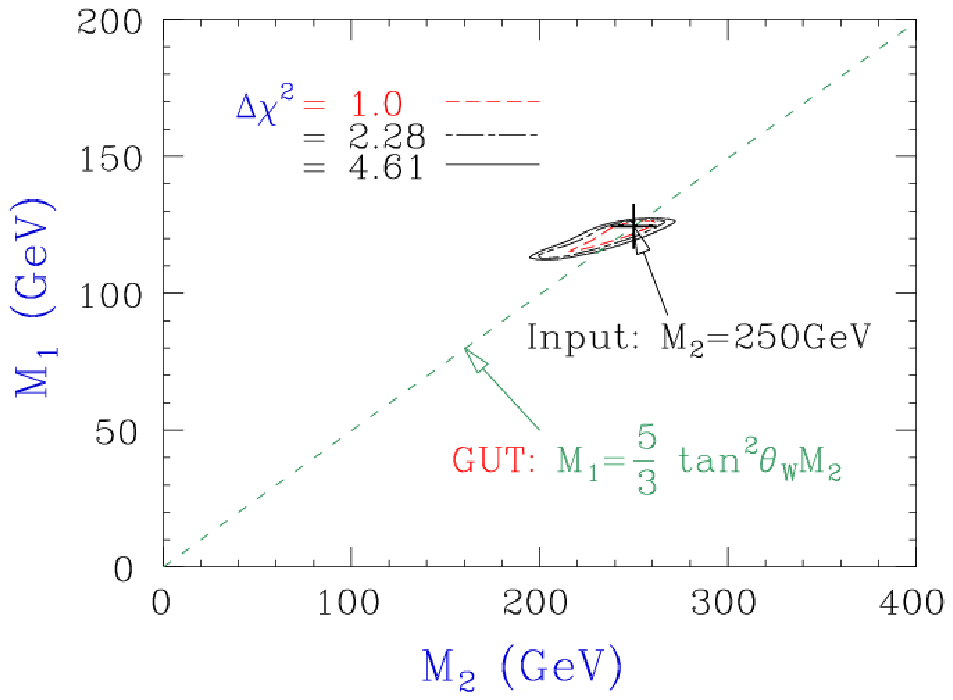}
\includegraphics[width=8cm,height=8cm]{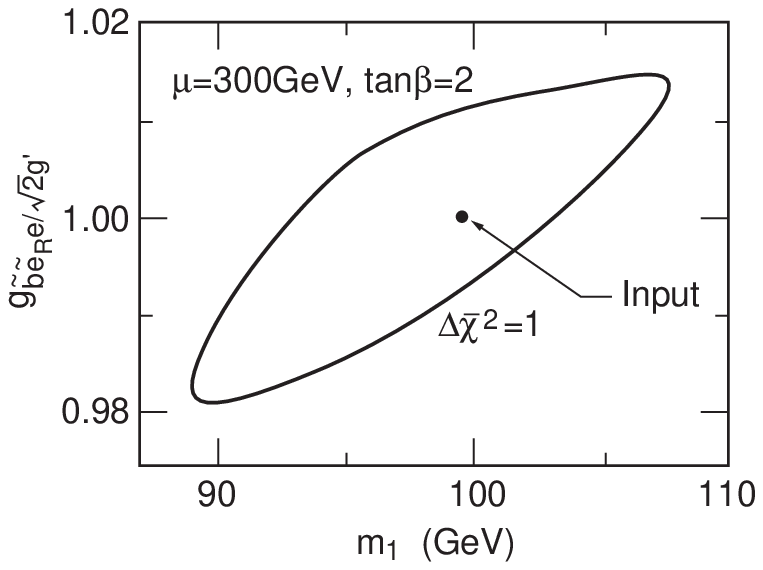}}
\caption{\label{lcsusytest}{\em Experimental confirmation at the
LC of the GUT relation between the gaugino masses $M_1$ and $M_1$.
b) Accuracy with which the SUSY gauge coupling can be measured at
the LC\/, from \cite{JapanSUSY}.}}
\end{center}
\end{figure*}

In this respect an \epemt machine with polarised beams is a
wonderful machine when it comes to providing model-independent
precision measurements\cite{JapanSUSY}. For instance, take the
issue about the nature of the slepton in the process of  pair
production of a right-handed smuon which most probably will decay
into the LSP neutralino and a muon. The signature is the same as
that of $W$ pair production with the $W$'s decaying into muons and
neutrinos and would constitute a formidable background. The use of
polarisation becomes almost a must. First of all, $W$ pair
production which is essentially an SU(2) weak process can be
switched off by choosing right-handed electrons. Indeed, at
high-energy one recovers the symmetric case where the $Z$ and
$\gamma$ separate into the orthogonal $W^0$ and $B$ (hypercharge).
The former not coupling to right-handed states. On the other hand
the same argument shows that if only the hypercharge boson is
exchanged and the fact that the hypercharge of the right-hand
electron is twice that of the left-handed one, right smuon
production will be four times larger than with left-handed $e^-$.
Thus polarization achieves three things: tags the nature of the
smuon (right-handed) independently of how it decays, increases the
signal cross section and dramatically decreases the background.
The smuon mass can be inferred either from a threshold scan which
is independent of the decay or as is the case here, the
measurement of the end-points of the muon energy which give both
the smuon mass and the LSP mass. A combined fit, for the case
above and for a modest luminosity $20fb^{-1}$, gives these masses
at the $1\%$ level. One more thing, to confirm the scalar nature
of the smuon one can look at its angular distribution which should
show a $sin ^2\theta$ dependence. In the case of the right-handed
selectron, this will not be the case since even with a
right-handed electron on has to deal with a t-channel neutralino
exchange. For the same reason as above only the bino component of
the neutralino will be selected. If this component is not
negligible one should observe a forward peak. This component is a
function of the gaugino parameters $M_{1,2}$, the $\mu$ parameter
and $tg \beta$. With the knowledge of $\chi^0_1$ one can measure
how much of the LSP is bino. Similar beautiful experiments with
chargino and neutralinos production can be conducted. They allow
to check the gaugino unification condition and allow also to
measure the couplings of the susy particle and verify them against
the gauge couplings, see Figs.~\ref{lcsusytest}.
Another important reason for extracting these model independent
parameters, is that they allow to calculate the relic density. It
could well be that the calculation turns out not to be compatible
with the required value. This would urge us to review some of our
assumptions concerning the nature of Dark Matter and/or the
hypotheses that enter these
calculations\cite{DarkmatterUnconventional}. Of course to be sure
to cover the parameter space as much as possible one needs to
access as many states as possible, and this is guaranteed only by
the highest energy linear collider.

\section{Conclusions}
The standard model has been extremely successful, but the jewel on
the crown, the Higgs is still missing. The latest LEP data shows
that this Higgs should be discovered at the LHC and extremely well
studied at a moderate energy linear collider that could be built
with existing technology (For a review of physics at the \epemt
see \cite{LCstudies}). Though theoretically unnatural, a scenario
with a heavy Higgs is still possible. However even this scenario
predicts New Physics at the TeV scale, but unfortunately without a
guarantee for direct observation at the LHC and a phase-I linear
collider. These kind of scenarios require probably a next
generation machines. The most motivated scenario is supersymmetry
which has a good chance to be, even though partially, discovered
at the LHC. However observation of supersymmetric particles is not
enough. One needs to understand the breaking of SUSY by measuring
as many of the parameters of the model as possible because this is
a probe into the physics of a hidden sector at unification of even
string scales. LHC could provide a few of these measurements but a
polarised \epemt machine is an ideal tool. In order to fully cover
the parameter space one may need a second phase of a ``leptonic"
machine at a few TeV. Of course other scenarios we still have not
thought of are possible. The recent hypothesis of extra-dimensions
should teach us to be cautious and not see the future through too
strait-laced arguments.

{\bf Acknowledgements}

I would like to thank Emilian Dudas for a clarification about
gauge unification at low scale, Genevi\`eve B\'elanger and Rohini
Godbole for reading the manuscript.

 \end{document}

%% file: wwww.tex
{ \unitlength=1.5 pt \SetScale{1.5}
\SetWidth{0.7}  }    
{} \qquad\allowbreak
\begin{picture}(79,65)(0,0)
\Text(13.0,57.0)[r]{$W^+$}
\DashArrowLine(14.0,57.0)(31.0,49.0){3.0}
\Text(13.0,41.0)[r]{$W^-$}
\DashArrowLine(31.0,49.0)(14.0,41.0){3.0}
\Text(39.0,50.0)[b]{$\gamma,Z$}
\DashLine(31.0,49.0)(48.0,49.0){3.0} \Text(66.0,57.0)[l]{$W^+$}
\DashArrowLine(48.0,49.0)(65.0,57.0){3.0}
\Text(66.0,41.0)[l]{$W^-$}
\DashArrowLine(65.0,41.0)(48.0,49.0){3.0}
\end{picture} \
{} \qquad\allowbreak
\begin{picture}(79,65)(0,0)
\Text(13.0,57.0)[r]{$W^+$}
\DashArrowLine(14.0,57.0)(48.0,57.0){3.0}
\Text(66.0,57.0)[l]{$W^+$}
\DashArrowLine(48.0,57.0)(65.0,57.0){3.0}
\Text(47.0,49.0)[r]{$\gamma,Z$}
\DashLine(48.0,57.0)(48.0,41.0){3.0} \Text(13.0,41.0)[r]{$W^-$}
\DashArrowLine(48.0,41.0)(14.0,41.0){3.0}
\Text(66.0,41.0)[l]{$W^-$}
\DashArrowLine(65.0,41.0)(48.0,41.0){3.0}
\end{picture} \
{} \qquad\allowbreak
\begin{picture}(79,65)(0,0)
\Text(13.0,65.0)[r]{$W^+$}
\DashArrowLine(14.0,65.0)(48.0,49.0){3.0}
\Text(13.0,33.0)[r]{$W^-$}
\DashArrowLine(48.0,49.0)(14.0,33.0){3.0}
\Text(66.0,57.0)[l]{$W^+$}
\DashArrowLine(48.0,49.0)(65.0,57.0){3.0}
\Text(66.0,41.0)[l]{$W^-$}
\DashArrowLine(65.0,41.0)(48.0,49.0){3.0}
\end{picture}
\vspace*{-1.5cm}

%% file: wwwwH.tex
{ \unitlength=1.5 pt \SetScale{1.5}
\SetWidth{0.7}   }   
{} \qquad\allowbreak
\begin{picture}(79,65)(0,0)
\Text(13.0,57.0)[r]{$W^+$}
\DashArrowLine(14.0,57.0)(31.0,49.0){3.0}
\Text(13.0,41.0)[r]{$W^-$}
\DashArrowLine(31.0,49.0)(14.0,41.0){3.0} \Text(39.0,50.0)[b]{$H$}
\DashLine(31.0,49.0)(48.0,49.0){3.0} \Text(66.0,57.0)[l]{$W^+$}
\DashArrowLine(48.0,49.0)(65.0,57.0){3.0}
\Text(66.0,41.0)[l]{$W^-$}
\DashArrowLine(65.0,41.0)(48.0,49.0){3.0}
\end{picture} \
{} \qquad\allowbreak
\begin{picture}(79,65)(0,0)
\Text(13.0,57.0)[r]{$W^+$}
\DashArrowLine(14.0,57.0)(48.0,57.0){3.0}
\Text(66.0,57.0)[l]{$W^+$}
\DashArrowLine(48.0,57.0)(65.0,57.0){3.0} \Text(47.0,49.0)[r]{$H$}
\DashLine(48.0,57.0)(48.0,41.0){3.0} \Text(13.0,41.0)[r]{$W^-$}
\DashArrowLine(48.0,41.0)(14.0,41.0){3.0}
\Text(66.0,41.0)[l]{$W^-$}
\DashArrowLine(65.0,41.0)(48.0,41.0){3.0}
\end{picture}
\vspace*{-1.5cm}
\vspace*{-1cm}